\documentclass[%
 reprint,
superscriptaddress,
 amsmath,amssymb,
 aps,
]{revtex4-2}

\usepackage{graphicx}
\usepackage{dcolumn}
\usepackage{bm}
\usepackage{color}
\usepackage{subfigure}

\begin{document}

\preprint{APS/123-QED}

\title{Polymeric Properties of Higher-Order G-Quadruplex Telomeric Structures: Effects of Chemically Inert Crowders}

\author{Deniz Mostarac}
\email{deniz.mostarac@uniroma1.it}
\affiliation{%
Department of Physics, University of Rome La Sapienza, 00185 Rome, Italy}

\author{Mattia Trapella}
\affiliation{%
 Department of Physics and Geology, University of Perugia, 06123 Perugia, Italy}%

\author{Luca Bertini}
\affiliation{%
 Department of Physics and Geology, University of Perugia, 06123 Perugia, Italy}%

\author{Lucia Comez}
\affiliation{%
Department of Physics and Geology, University of Perugia, 06123 Perugia, Italy}%
 
\author{Alessandro Paciaroni}
\affiliation{%
 Department of Physics and Geology, University of Perugia, 06123 Perugia, Italy}%

\author{Cristiano De Michele}
\affiliation{%
Department of Physics, University of Rome La Sapienza, 00185 Rome, Italy}%

\date{\today}
\begin{abstract}
G-quadruplexes are non-canonical DNA structures rather ubiquitous in human genome, which are thought to play a crucial role in the development of 85-90 \% of cancers.
Here, we present a novel coarse-grained approach in modeling G-quadruplexes which accounts for their structural flexibility. We apply it to study the polymeric properties of G-quadruplex multimers, with and without crowder particles, to mimic in-vivo conditions. We find that, contrary to some suggestions found in the literature, long G-quadruplex multimers are rather flexible polymeric macromolecules, with a local persistence length comparable to monomer size, exhibiting chain stiffness variation profile consistent with a real polymer in good solvent. Moreover, in a crowded environment (up to 10$\%$ volume fraction), we report that G-quadruplex multimers exhibit an increased propensity for coiling, with a corresponding decrease in the measured chain stiffness. Accurately accounting for the polymeric properties of G4 multimers is crucial for understanding their interactions with anticancer G4-targeting drugs, thereby significantly enhancing the design and effectiveness of these drugs.
\end{abstract}

\maketitle


G-quadruplexes (G4s) are non-canonical DNA conformations, formed by guanine-rich oligonucleotides. Structurally, a G4 consists of an array of quasi-planar tetrads of guanine tracts (G-tetrads). In the telomeric overhangs of human chromosomes a G4 usually contains three G-tetrads.  Although G4 structures are said to occur in three main topologies~\cite{webba2007geometric,neidle2003structure,karsisiotis2011topological} (parallel, anti-parallel, and hybrid), it is widely recognized that they are highly polymorphic structures~\cite{lim2009structure,lim2013structure,wang1993solution,parkinson2002crystal,ambrus2006human,dai2007structure,dai2007structure2,luu2006structure,phan2007structure,zhang2010structure,phan2006different,li2005not,dai2008polymorphism} with long folding time scales, long-living quasi-stable conformations, and common coexistence of multiple G4 topologies in solution~\cite{long2013kinetic,gray2014folding,you2014dynamics,bessi2015involvement,armstrong2015nanometal,xue2011kinetic,lannan2012human,gabelica2014pilgrim}. Regardless of the specific oligonucleotide sequence, the morphology of a G4 is largely achieved via a network of Hoogsteen-type hydrogen bonds, pi-stacking interactions, and coordinating cations~\cite{lane2008stability,biffi2013quantitative}, and is contingent on environmental factors such as the cation type and concentration, molecular crowding, and dehydration conditions~\cite{huppert2010structure,phan2006dna,viglasky2011first,chaires2010human,smargiasso2008g,heddi2011structure}.

Biological role(s) of G4 DNA and its metabolizing enzymes (e.g., helicases) in DNA transcription and genomic stability are not fully understood yet. Sequences capable of forming G4s are abundant in the genomes of higher eukaryotes~\cite{huppert2005prevalence,todd2005highly,hansel2017dna}. Such sequences are particularly concentrated in telomeric regions, constituting up to 25$\%$ of all DNA G4s~\cite{biffi2013quantitative}. G4s have been observed in vivo~\cite{schaffitzel2001vitro,henderson2013detection,bao2019hybrid,lam2013g}, where they are believed to play a role in regulating transcription, translation, DNA replication, RNA localization, and various other crucial biological functions~\cite{huppert2007g,kendrick2010role,rhodes2015g}. Due to their significance in biology, G4s have received considerable attention as targets for drug design~\cite{bianchi2018structure,comez2020polymorphism}. Since G4s have been shown to inhibit telomerase and HIV integrase~\cite{zahler1991inhibition}, there is potential for specific G4-stabilizing compounds to be utilized as anticancer or antiviral medications~\cite{tahara2006g,zhou2016telomere,neidle2010human,perrone2014anti}. Moreover, G4s have been extensively explored as promising building blocks in synthetic biology and nanotechnology~\cite{yatsunyk2014nano,mergny2019dna}.

While much of the research on G4s has concentrated on their monomeric state, G4s occur in various multimeric configurations~\cite{kolesnikova2019structure}, with distinct biological roles and special interest as potential drug targets~\cite{frasson2022multimeric,rigo2022polymorphic}. By a G4 multimer, it is implied that for a G-tract array, distinct interfaces that delimit repeating loop motifs between groups of successive G-tracts can be noted. These transient stacking interfaces could be viewed as binding grooves, which may be valuable for drug targeting. Given that single-stranded telomeric overhang length ranges from 50 to $>600$ nucleotides~\cite{rk1988highly,wright1997normal}, with a conservative estimate of the number of nucleotides needed for a G4 to form being $\approx 25$, it is not surprising that higher-order structures such as G4 multimers form~\cite{monsen2022g}. G4 multimers tend to form in biological environments that are densely packed with various biomolecules~\cite{bajpai2020mesoscale}. It has been reported that crowder molecules (CMs) tend to stabilize G4s and support the formation of higher-order structures~\cite{aznauryan2023dynamics,heddi2011structure,gao2023effects}.

However, there is some disagreement about the telomeric higher-order structure formation in solution. Some literature suggests maximal G-tract usage in extended human telomere G-rich sequences when forming G-quadruplex structures that are describable as beads-on-a-string~\cite{yu2006characterization,renvciuk2009arrangements,xu2009consecutive,abraham2014interaction,bugaut2015understanding}. Others agree on G-tract utilization but propose a more rigid structure, with extended telomere G4s adopting compact, somewhat rod-like structures via stacking interactions~\cite{petraccone2010integrated,petraccone2011structure,chaires2015hydrodynamic}. On the other hand, some publications report highly flexible arrangements with large gaps occurring between G-quadruplexes, indicating that G-tract utilization is not maximized~\cite{wang2011single,kar2018long,abraham2018random}.

Consequently, there is currently no clear view on the flexibility of G4 multimers. This is a crucial question to answer, as flexibility relates to the functions of biopolymers~\cite{henzler2007dynamic,karplus2005molecular,zimmerberg2006proteins,pyle2002metal}, and needs to be quantified in order to be able to scrutinize any physical quantities that change in space according to the distance with respect to the object of interest~\cite{hsu2009define}. In particular, the ability of anticancer ligands (i.e. drugs such as BRACO-19, TMPyP4, etc.) to stabilize G4s could strongly affect multimer flexibility~\cite{zhao2020recent,frasson2022multimeric,figueiredo2024g}.

Counterion concentration, distribution, and the phenomenon of counterion condensation is a typical example for polyelectrolyte chains like, e.g., DNA~\cite{manning1969limiting}. This is especially relevant for higher-order G4 structures, as researchers try to understand their complex interplay with various ligands, drugs, and crowding molecules.

The matter is complicated by the fact that there is little structural information on multimeric G4s, as X-ray crystallography and/or Nuclear Magnetic Resonance spectroscopy studies have not been able to deal with longer nucleic acid sequences~\cite{monsen2022g}. Much of the discrepancies found in the literature could be associated with the inherent limitations of present experimental techniques (e.g., spatial resolution or sample tampering via dehydration). To that end, recent advancements in small-angle X-ray scattering (SAXS) experiments have significantly enhanced our understanding of the higher-order structures telomeric sequences form in solution. However, extracting the relevant information from SAXS patterns necessitates the use of complex ab initio space-filling models or atomistic simulations~\cite{monsen2021solution,limongelli2013g,galindo2016assessing,islam2013conformational,wei2015flexibility,ghosh2015plant,bian2014atomistic,yang2017silico,stadlbauer2019parallel,rocca2020folding,pokorna2024molecular,kim2012free,bergues2015role,zeng2016unfolding,luo2016computational,kogut2016molecular,gajarský2017structure,bian2018exploration,bian2020insights,stadlbauer2021insights}. Despite the use of various enhanced sampling methods, even in combination with united-atom representations, the extreme computational cost, system size, and timescale restrictions inherent to atomistic simulations limit their utility in the study of higher-order G4 structures. Furthermore, strength of stacking interactions between G4s, which is crucial to determine G4 multimers conformation, is not well reproduced by current atomistic force fields~\cite{Maffeo2012,Maffeo2014,GlaserPRE17}.

Recently, using extremely coarse-grained Monte Carlo simulations, we enabled the direct interpretation of in vitro SAXS experiments on the self-assembly of Tel22 (d(TTAGGG)3)) and Tel72 (d(TTAGGG)12) multimers, with and without ligands (TMPyP4 porphyrin and BRACO-19, respectively)~\cite{rosi2023stacking}. However, this approach cannot be used to scrutinize phenomenology where resolving the structural features of G4s is necessary (length scale less than a few nanometers). Here, we present a semi-atomistic coarse-grained model of G4 mono- and multimers, validated against in-vitro experimental data. Using the Espresso simulation package~\cite{weik2019espresso}, we perform long timescale, bulk Langevin Dynamics~\cite{allen2017computer} simulations of G4 multimers, $M \times G4$, where $M$ denotes the number of monomers and $M \in {1, 2, 3, 4, 10, 15, 20}$, to scrutinize the polymeric properties of higher-order G4 structures, also in the presence of crowding molecules. This contribution provides a deep insight into long telomeric G4 multimers, and sets expectations for future in-vitro studies.

\begin{figure}[h]
    \centering
    \includegraphics[width=\linewidth]{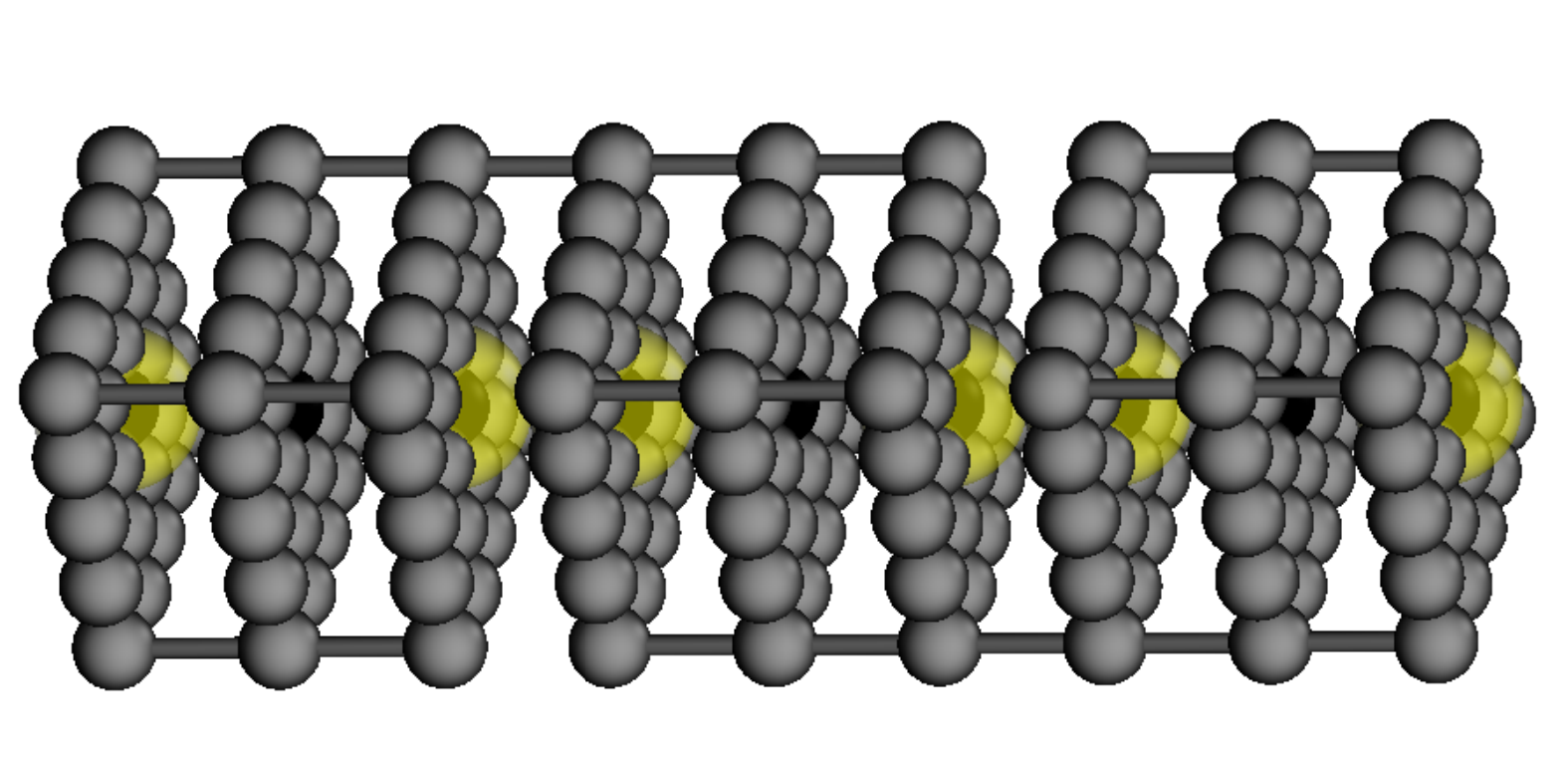}
    \caption{Schematic representation of the coarse-grained representation of a G4 trimer. The CoM particles are colored black. Particles outlining a G-tetrad and the links between adjacent G4 monomers are coloured gray. The central attraction between the CoM particles on the outer G-tetrads are depicted as transparent yellow spheres. Relative sizes and distances correspond to the interaction minima, respectively.}
    \label{fig:quad_doodle}
\end{figure}

The baseline structure in our simulations is the G-tetrad, modeled as a $5\times 5$ grid of equidistant soft spheres~\cite{weeks1971role}, with a characteristic diameter $\sigma$, making out a thin square sheet. A G4 monomer consists of three G-tetrads, linked together via FENE bonds~\cite{kremer1990molecular}. Specifically, the corner particles in adjacent G-tetrads in a G4 are linked. Making multimeric structures out of G4 monomers is achieved by introducing FENE linkers between a randomly chosen pair of corner particles on adjacent G-tetrads of neighbouring G4 monomers. In order to mimic the stacking interactions between monomers~\cite{rosi2023stacking}, the center-of-mass (CoM) particles of the outer G-tetrads exhibit a central attraction, realised via Lennard-Jones interaction potential. Details about the simulations approach, model and units are provided in the Supplementary Information.

This model is deliberately minimalist, with the purpose to minimise complexity, thus maximing efficiency. It also reflects a particular view on the structure of a G4. Consider a single telomeric G4 monomer, which is folded out of a AG3(T2AG3)3~\cite{libera2021porphyrin} or 2JSL sequence~\cite{monsen2021solution}. In physiological conditions, such a monomer consists of necessarily three G-tetrads that contain two $K+$ or three $Na+$ stabilizing cations. In fact, most of the structural stability of a G4 monomer comes from the electrostatic interaction (in this context the hydrogen bonds are also electrostatic interactions) between the G-tetrads and the ions within the G4~\cite{yu2006characterization,bugaut2015understanding,petraccone2011structure}. Given that, the electrostatic interactions within the G4 are effectively short-ranged due to evident interaction screening, we take the view that a G-tetrad can be represented as a purely topological, steric hindrance. Due to the short range, but strong electrostatic interaction, each G-tetrad is considered as firmly coupled to a monovalent ion. Since the G-tetrads in a G4 monomer are linked via short but elastic liners, whereas the inter-monomer links are comparable to the average inter-tetrad links, the overall structure is rather soft. A depiction of this view on the structure of G4 monomers and multimers can be seen in Fig.~\ref{fig:quad_doodle}

\begin{figure}[ht]
    \centering
\label{fig:tel48}\includegraphics[width=\linewidth]{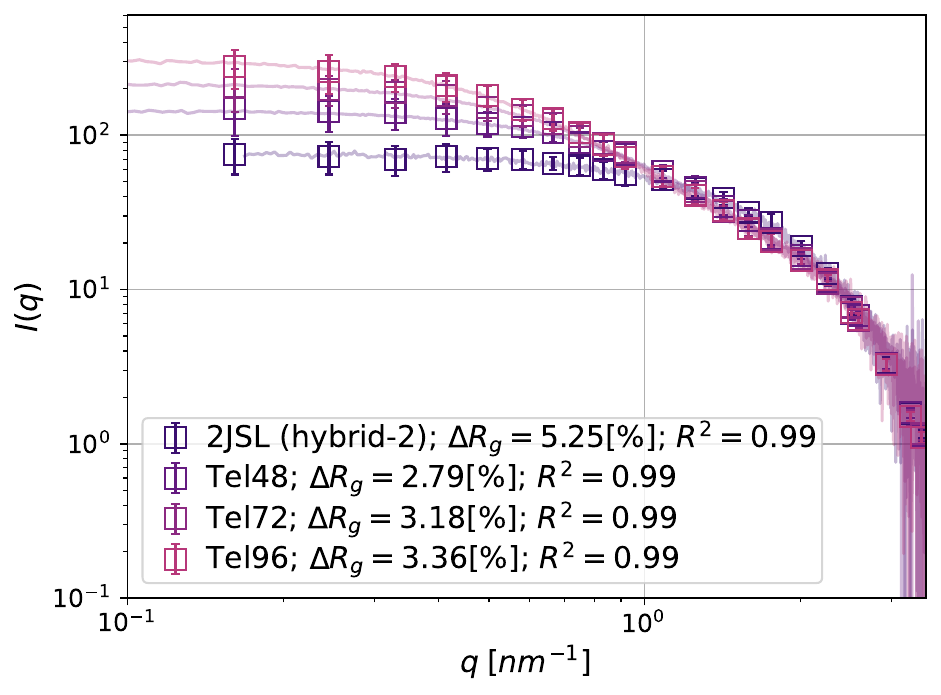}
    \caption{Scattering intensity $I(q)$ comparison between in-vitro G4 multimers published in \citet{monsen2021solution} and coarse-grained simulations. The fit between the simulated and experimental data was achieved by determining two scaling factors ($q$ and $I(q)$ axis, respectively) that maximise the coefficient of determination, $R^2$ (shown in the legend). Each subplot is showing a measure of relative error in the extracted value of $R_g$ from Guinier analysis, between simulated and 
    experimental results, $\Delta R_g$ (fit parameters provided in Supplementary Information). Error-bars are showing the standard deviation of simulation data.}
    \label{fig:master_mag}
\end{figure}
We tuned and validated our model by comparing simulated results with the experimental results reported by \citet{monsen2021solution}. The experimental study in question, reports in-vitro experiments using the 2JSL, Tel48, Tel72, and Tel96 sequences. In our study, these correspond to $M\times G4$ multimers where $M \in \{1, 2, 3, 4\}$, respectively.

Looking at Fig.~\ref{fig:master_mag}, one can see that our coarse-grained approach captures the experimental SAXS data very well. For an in-depth discussion on the experimental systems, we refer the reader to the exhaustive analysis presented in the original paper. Here, we only summarise the key points relevant for the purposes of this contribution. Our simulations corroborate that in the in-vitro experiments, G4 multimers effectively do not  interact between each other. This is true in our simulations by construction. The differences in curvature of the intermediate to high-$q$ range can entirely be attributed to increasing monomer number between the samples, which supports the claim that the samples maximize the possible monomer number. In the low-$q$ range, the scattering intensity $I(q)$ can be approximated as:
\begin{equation}
    I(q)\approx I(0) e^{-q^2 R_g^2 /3}.
\end{equation}
This is the well known Guinier’s approximation~\cite{guinier1955small}, which is often used in experimental studies to extract the radius of gyration $R_g$~\cite{feigin1987structure}. Formally, the radius of gyration is defined as
\begin{equation}\label{eq:rg_direct}
    R_g = \sqrt{\lambda_1^2+\lambda_2^2+\lambda_3^2},
\end{equation}
where $\lambda_1>\lambda_2>\lambda_3$ are the eigenvalues of the gyration tensor:
\begin{align}
    G_{\mu \nu}=\dfrac{1}{N}\sum^N_{i=1}(r_{i,\mu}-r_{cm,\mu})(r_{i,\nu}-r_{cm,\nu}),
\end{align}
where $r_{i,\mu}$ and $r_{cm,\mu}$ are the $\mu$-th Cartesian components of the position of the $i$-th particle and the center of mass, respectively. The summation is carried over all $N$ particles. We define $\Delta R_g=|R^{sim}_g-R^{exp}_g|/R^{exp}_g$ as a measure of the difference between the radii of gyration of the experimental systems and their coarse-grained representations using our model, $R^{exp}_g$ and $R^{sim}_g$, respectively, extracted via the Guinier analysis. We define $\delta R_g=|R^{guinier}_g-R^{direct}_g|/R^{direct}_g$ as a measure of the difference between the radii of gyration measured from simulated data via Guinier analysis and directly from the data using Eq.~\ref{eq:rg_direct}, $R^{guinier}_g$ and $R^{direct}_g$, respectively.  We obtained $\Delta R_g$ values of a few percent, indicating a very good agreement. On the other hand, as can be seen in Table~\ref{tbl:rg}, $\delta R_g$ indicates that there can be up to a $10\%$ discrepancy in measured $R_g$ depending on the measurement approach. In our view, extracting $R_g$ from computational representations of the experimental system that reproduce the experimental measurements is a transparent and preferable approach compared to Guinier analysis. This highlights the role of highly scalable, coarse-grained models, such as the one presented here, as tools where a reduced set of fit parameters (reduced complexity) can be tuned to match experimental measurements and, through that, enable further insights.
\begin{table}[h!]
    \centering
    \begin{tabular}{|c|c|c|c|c|}
        \hline
        & $1\times G4$ & $2\times G4$ & $3\times G4$ & $4\times G4$ \\ 
        \hline
        $\delta R_g [\%]$ & 2.9377 & 4.1977 & 8.5696 & 6.0594 \\ 
        \hline
    \end{tabular}
    \caption{Summarized results showing percent difference in measured radius of gyration, $\delta R_g$, based on measurement approach, where we used either Guinier analysis or direct analysis using Eq.\ref{eq:rg_direct}}
    \label{tbl:rg}
\end{table}

In \citet{monsen2021solution}, it is stated that the flexibility of G4 multimers are semi-flexible polymers, consistent with rigid G4 units linked by flexible, hinged, interfaces. Similar statements (and contradictory ones) can be found across the literature summarised above, where $R_g$ as a function of monomer numbers is fitted with a random Gaussian coil and/or the Worm-like chain model to estimate persistence length $L_p$. These models are known to reproduce the stiffness of canonical duplex DNA~\cite{bustamante1994entropic}. While such an analysis is certainly useful, it is not sufficient to characterise the flexibility of G4 multimers. In order to be able to do so, it is necessary to discuss longer $M\times G4$ multimers where $M \in \{4, 10, 15, 20\}$. 
\begin{figure}[h]
    \centering
    \includegraphics[width=\linewidth]{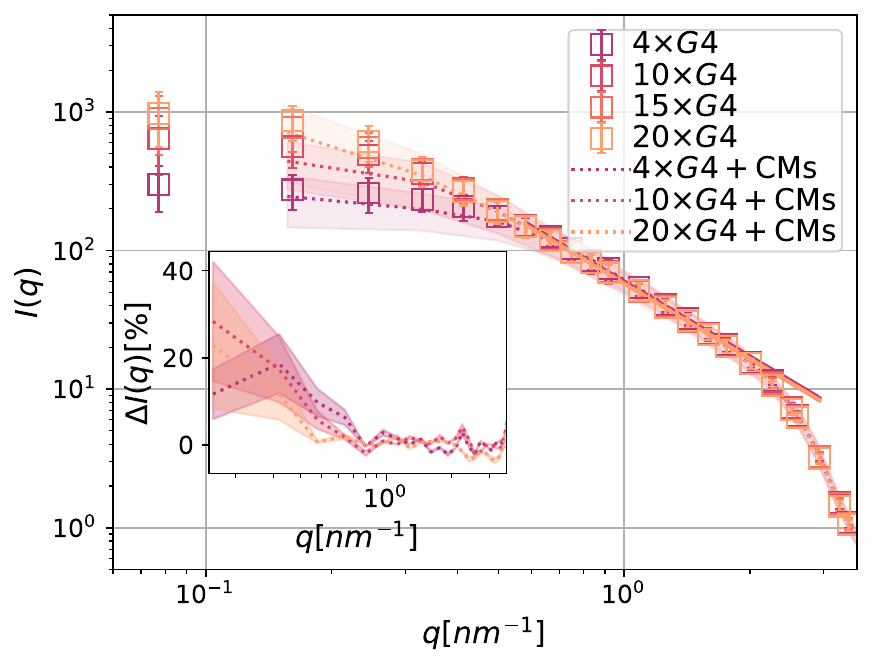}
    \caption{Simulated scattering intensities $I(q)$ for long $M \times G4$ multimers. We fitted the slope of the intermediate-$q$ range plateau (linear in a log-log plot) using $f(q)=a+bq$, where we obtained $b=-1.82 \pm 0.01$. Data without CMs are shown with square symbols, and simulations were performed at $C=0.6$ mM. Error bars represent the standard deviation of the simulation data. Corresponding data from simulations with CMs are displayed with equi-colored dotted lines, obtained via cubic spline interpolation for clarity. These simulations were performed at $C=5$ mM. Confidence intervals are provided. The inset shows the percentage difference in the measured scattering intensity, $\Delta I(q)=100 \times |I(q)/I_{\text{CM}}(q)-1|$, between simulations with and without CMs ($I_{\text{CM}}(q)$ and $I(q)$, respectively). The inset axes are shown in linear-log scale, and confidence intervals are calculated via standard error propagation.}
    \label{fig:sq_long}
\end{figure}

Looking at the $I(q)$ for long G4 multimers, shown in Fig.~\ref{fig:sq_long}, one notes the formation of two plateaus, in the low and the intermediate-$q$ range, respectively. In the low-$q$ range we see the asymptotic approach to a maximal $I(q)$ height, with increasing G4 monomer number. Furthermore, the scattering profiles do not approach the $y$-axis horizontally for $15\times G4$ and $20\times G4$ multimers, which signals inter-particle interactions (i.e. multimer self-interactions as backbone crossing is not allowed). Both of these points are consistent with an image of flexible, coiling polymer. The slope of the $I(q)$ plateau in the intermediate-$q$ range can be related with the distribution of bond vectors in a polymeric sample, where it is known that a slope of exactly $-2$ corresponds to a Gaussian distribution~\cite{beaucage1996small,beaucage1995approximations}. The slope we extract, however, hints that there are non-trivial inter-monomer correlations along the polymeric backbone.

As previously stated, G4 multimers tend to form in biological environments that are densely packed with various biomolecules and potentially exhibit complex chemical interactions with them. In the first approximation, however, the primary modifier of their in-vitro measured properties, such as the polymeric properties, is due to high proportion of excluded volume. To study the effect of excluded volume a crowded environment on the properties of G4 multimers, we simulated $M\times G4$ multimers where $M \in \{4,10,15,20\}$ at $1\%$ and $10\%$ volume fraction of CMs. The CMs are represented as soft spheres with a interaction range $\sigma_{crowder}=6$, which is comparable to the long diagonal of a circumscribed sphere for a single G4 monomer in our simulations. This simulation setup is aimed to set expectations for in-vitro studies of crowded G4 multimers in a good solvent, suspended with long $PEG$ molecules for example~\cite{aznauryan2023dynamics}. Looking at Fig.~\ref{fig:sq_long}, where we also provide the scattering profiles for $4\times G4$, $10\times G4$ and $20\times G4$ multimers with the CMs at volume fraction $\phi=10\%$, we can see that the presence of crowding molecules can be noted as a reduction in the $I(q)$ in the low-$q$ region, with a correspondingly increased variance. As can be seen from the inset in Fig.~\ref{fig:sq_long}, the noted reduction in the $I(q)$ is statistically significant, and can be understood as increased coiling propensity of G4 multimers in a crowded environment. We could
expect that this propensity would be largely enhanced in the much more crowded environment of human cells -- where volume fraction of CMs is estimated to be around 30-40\% ~\cite{Sharp2016} -- thus becoming an important aspect to take into account in the development of G4 targeting ligands.

With this, we turn our attention to the flexibility of long G4 multimers. Commonly, flexibility of macromolecules is characterised using the notion of persistence length~\cite{2003-rubinstein}. Persistence length is a well defined quantity only for ideal chains following Gaussian statistics and its calculations are contingent in strong theoretical assumptions. Classically, $L^{id}_p$ is calculated from the decay of the auto-correlation function 
\begin{equation}\label{eq::cn_classic}
    C(N_b)=<cos\theta_{k,k+N_b}>=<\vec{a_k}\cdot \vec{a}_{k+N_b}>
\end{equation}
between vectors $\vec{a}_k$ connecting each pair $k$ of neighbouring monomers along the backbone, separated by $N_b$ bond vectors, where a bond vector is defined as the center-of-mass distance between a pair of adjacent monomers. The expectation is that
\begin{equation}\label{eq::lp_classic}
    C(N_b)\approx \exp{\left (-\frac{N_bL_{b}}{L^{id}_p}\right )},
\end{equation}
where $L_{b}$ is the average bond vector length. This expectation is justified only for a Gaussian distribution of bond vector orientations. For real polymers, this is not the case, as non-trivial excluded volume correlations persist throughout the polymeric backbone, and exhibit a reported power law decorrelation profile~\cite{hsu2010standard}. More generally, persistence length is a chain property that can, within real polymer theory, vary substantially along the chain backbone, and can be defined as:
\begin{equation}\label{eq::lp_real}
    \dfrac{L^{re}_p(k)}{L_b}=<\vec{a_k}\cdot \vec{R_e}/|\vec{a_k}|^2>,
\end{equation}
where $R_e$ is the end-to-end distance vector. \citet{schafer2004calculation} have shown that, to a very good approximation:
\begin{equation}\label{eq::schafer}
    \dfrac{L^{re}_p(k)}{L_b}\approx\alpha \Bigg[\dfrac{k[(M-1)-k]}{M-1}\Bigg]^{2\nu -1}.
\end{equation}

\begin{figure}[ht]
    \centering
    \includegraphics[width=\linewidth]{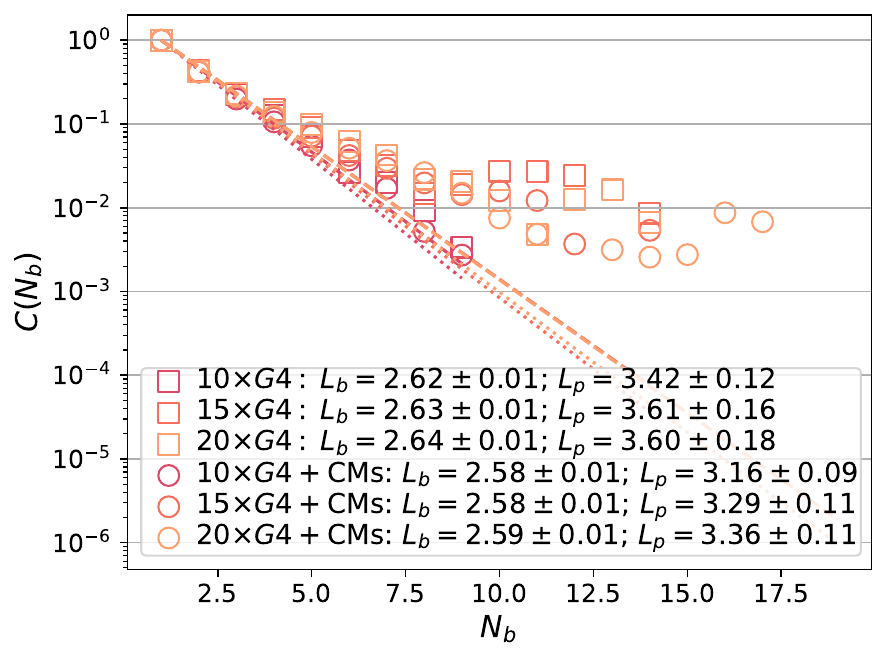}
    \caption{Bond correlation function $C(N_b)$, as defined by Eq.~\ref{eq::cn_classic}. Data for G4 multimer simulations without CMs are shown with square symbols, while data with 10$\%$ CMs volume fraction are shown with circles. The fit of Eq.~\ref{eq::lp_classic} is represented by dash-dotted lines for simulations with CMs and dashed lines for those without CMs. Extracted $L_p$ values are shown in the legend. The y-axis is presented on a logarithmic scale. Color coding is provided in the legend.}
    \label{fig:classic_lp}
\end{figure}

\begin{figure}[ht]
    \centering
    \includegraphics[width=\linewidth]{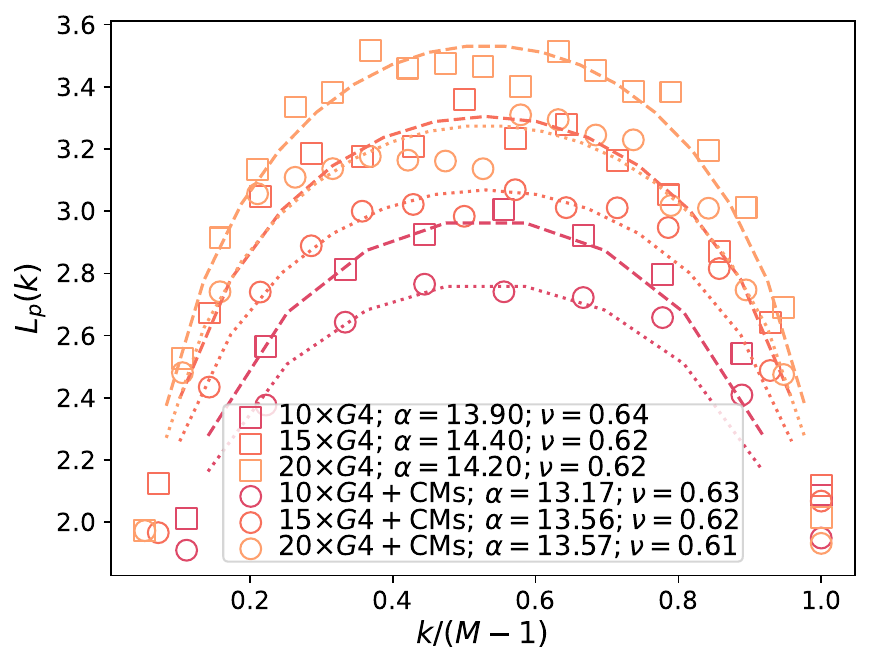}
    \caption{Persistence length  $L_p$, defined by Eq.\ref{eq::lp_real}. Data for G4 multimer simulations without CMs are shown with square symbols, while data with 10$\%$ CMs volume fraction are shown with circles. Fits of Eq.\ref{eq::schafer} to the data are represented by dashed lines (no CMs) and dotted lines (CMs) in corresponding colors. The color coding is provided in the legend. The y-axis is presented on a logarithmic scale. }
    \label{fig:real_lp}
\end{figure}

Looking at Fig.~\ref{fig:classic_lp}, we can see how well the classic notion $L^{id}_p$ can be applied in the context of G4 multimers. The $C(N_b)$ seems to correspond to the expected exponential decay only for $10\times G4$ multimers. For longer multimers ($15\times G4$ and $20\times G4$), we observe systematic deviations from the exponential decay, signaling the onset of the a power law decay, characteristic of real polymers. It is important to note that the $M\times G4$ multimers we studied are relatively short from the perspective of polymer physics scaling theories~\cite{hsu2013estimation}. The deviations start being notable only for small correlations $C(N_b)\approx 0.05$. Therefore, we can use the $L^{id}_p$ values we extract and present their average as a monomer number independent estimate of the stiffness of G4 multimers, where we obtain $L^{id}_p=3.51\pm 0.01$, which is compatible with the values reported by \citet{monsen2021solution}. However, we have also seen clear indications that G4 multimers do not follow ideal polymer statistics. As such, while $L^{id}_p$ is a useful \textit{relative} quantity~\cite{forero2019backbone,micka1996persistence}, particularly from the point of view of experimental studies, it is not strictly correct to apply it to G4 multimers. Looking at Fig.~\ref{fig:real_lp}, we can see the $L^{re}_P$ fits to our simulated data on $10\times G4$, $15\times G4$ and $20\times G4$ well. This elucidates key properties to be expected from long G4 multimers, which is that, as is characteristic of real polymers, chain stiffness varies within a G4 multimer, well captured by the concave shape of Eq.~\ref{eq::schafer}. The $L^{re}_P$ and $L^{id}_P$ (monomer number independent) values we measured are compatible. Consistent with what we have observed in Fig.\ref{fig:sq_long}, the presence of CMs systematically decreases the $L_p$. Obtaining a measure of $L^{re}_P$ that is $M$ independent is not feasible as it is necessary to study much larger monomer numbers to make such an estimate sensible~\cite{hsu2010standard}. The matter is further complicated by the fact that it is highly unlikely for G4 multimers with a higher monomer number than we have studied here to form~\cite{wright1997normal,moyzis1988highly}. Having said that, the analysis we presented here is sufficient to show that, the scaling exponent is closer to the expected scaling factor for a real polymer in a good solvent~\cite{2003-rubinstein}. Our results underline that long G4 multimers behave as flexible polymers with a local chain stiffness comparable to monomer size, exhibiting properties consistent with a real polymer in a good solvent, in in-vitro conditions reported in \citet{monsen2021solution}. Our results support a view where stacking interactions between the monomers in a G4 multimer are rather weak. In this case, provided the short hinged interfaces between the monomers, it is clear that G4 monomers bend and twist away form each other to maximise entropy, in which case the steric hindrance coming form the monomer shape is not relevant. Preventing neighbouring G4 monomers to twist away form each other will require significant solvent induced hydrophobic interactions, at which point G4 multimers would probably also start to aggregate.
One of the most distinctive features of G4s, within the scope of canonical and non-canonical conformations DNA can assume, is the quasi-cubic monomer shape. Even for soft systems, monomer shape reflects on to polymeric properties substantially, provided that both the translational and rotational degrees of freedom between the monomers are coupled and the average inter-monomer distance is low enough~\cite{mostarac2022nanopolymers}. This does not turn out to be the case for G4 multimers, which can exhibit polymeric properties in line with a flexible real polymer in a good solvent, at least as far as we can see from in-vitro studies. It is a very interesting point to consider that the fact G4 multimers sit somewhat between single-strand DNA and duplex DNA in terms of flexibility~\cite{monsen2022g}, serves a functional purpose. It has been shown that ligands such as TMPyP4 porphyrin, broadly speaking, stack between G4 units~\cite{libera2021porphyrin} (yellow terminals in Fig.\ref{fig:quad_doodle}) and their action can be represented as an effective increase in stacking interaction strength~\cite{rosi2023stacking}. Moreover, this selective action provides a significant advantage in the use of G4 stabilizers as anticancer drugs~\cite{criscuolo2022insights,zegers2023dna}. In light of the results we presented here it is clear that such ligands increase the stiffness of G4 structures, that are otherwise entirely flexible, especially in a crowded complex biological environment. Therefore, we suggest that  the efficacy of anticancer G4 targeting ligands is closely related to the multimer stiffness increase they cause. Hopefully, this letter inspires further experimental studies to scrutinise this point. From the point of view of chemistry, systems involving G4 are tremendously complicated. However, for equilibrium properties of G4 multimers, such details can be omitted. The fact of the matter is that scalable coarse-grained models can and should be used to study G4 systems, unlocking a variety of implicit and explicit solvent simulation studies that were previously simply not feasible. The CG model we present here specifically, can be used to scrutinise dynamics of G4 systems, which is something we are currently working on. In this respect, this model could also allow to efficiently study the folding/unfolding pathways and aggregation kinetics of ligands and G4s.


All authors acknowledge financial support from European Union –
Next Generation EU (MUR-PRIN2022 TAMeQUAD CUP:B53
D23004500006). Computer simulations were performed at the Vienna Scientific Cluster (VSC-5).

\bibliography{quadriplex} 

\begin{thebibliography}{124}%
\makeatletter
\providecommand \@ifxundefined [1]{%
 \@ifx{#1\undefined}
}%
\providecommand \@ifnum [1]{%
 \ifnum #1\expandafter \@firstoftwo
 \else \expandafter \@secondoftwo
 \fi
}%
\providecommand \@ifx [1]{%
 \ifx #1\expandafter \@firstoftwo
 \else \expandafter \@secondoftwo
 \fi
}%
\providecommand \natexlab [1]{#1}%
\providecommand \enquote  [1]{``#1''}%
\providecommand \bibnamefont  [1]{#1}%
\providecommand \bibfnamefont [1]{#1}%
\providecommand \citenamefont [1]{#1}%
\providecommand \href@noop [0]{\@secondoftwo}%
\providecommand \href [0]{\begingroup \@sanitize@url \@href}%
\providecommand \@href[1]{\@@startlink{#1}\@@href}%
\providecommand \@@href[1]{\endgroup#1\@@endlink}%
\providecommand \@sanitize@url [0]{\catcode `\\12\catcode `\$12\catcode `\&12\catcode `\#12\catcode `\^12\catcode `\_12\catcode `\%12\relax}%
\providecommand \@@startlink[1]{}%
\providecommand \@@endlink[0]{}%
\providecommand \url  [0]{\begingroup\@sanitize@url \@url }%
\providecommand \@url [1]{\endgroup\@href {#1}{\urlprefix }}%
\providecommand \urlprefix  [0]{URL }%
\providecommand \Eprint [0]{\href }%
\providecommand \doibase [0]{https://doi.org/}%
\providecommand \selectlanguage [0]{\@gobble}%
\providecommand \bibinfo  [0]{\@secondoftwo}%
\providecommand \bibfield  [0]{\@secondoftwo}%
\providecommand \translation [1]{[#1]}%
\providecommand \BibitemOpen [0]{}%
\providecommand \bibitemStop [0]{}%
\providecommand \bibitemNoStop [0]{.\EOS\space}%
\providecommand \EOS [0]{\spacefactor3000\relax}%
\providecommand \BibitemShut  [1]{\csname bibitem#1\endcsname}%
\let\auto@bib@innerbib\@empty
\bibitem [{\citenamefont {Webba~da Silva}(2007)}]{webba2007geometric}%
  \BibitemOpen
  \bibfield  {author} {\bibinfo {author} {\bibfnamefont {M.}~\bibnamefont {Webba~da Silva}},\ }\bibfield  {title} {\bibinfo {title} {Geometric formalism for dna quadruplex folding},\ }\href@noop {} {\bibfield  {journal} {\bibinfo  {journal} {Chemistry--A European Journal}\ }\textbf {\bibinfo {volume} {13}},\ \bibinfo {pages} {9738} (\bibinfo {year} {2007})}\BibitemShut {NoStop}%
\bibitem [{\citenamefont {Neidle}\ and\ \citenamefont {Parkinson}(2003)}]{neidle2003structure}%
  \BibitemOpen
  \bibfield  {author} {\bibinfo {author} {\bibfnamefont {S.}~\bibnamefont {Neidle}}\ and\ \bibinfo {author} {\bibfnamefont {G.~N.}\ \bibnamefont {Parkinson}},\ }\bibfield  {title} {\bibinfo {title} {The structure of telomeric dna},\ }\href@noop {} {\bibfield  {journal} {\bibinfo  {journal} {Current opinion in structural biology}\ }\textbf {\bibinfo {volume} {13}},\ \bibinfo {pages} {275} (\bibinfo {year} {2003})}\BibitemShut {NoStop}%
\bibitem [{\citenamefont {Karsisiotis}\ \emph {et~al.}(2011)\citenamefont {Karsisiotis}, \citenamefont {Hessari}, \citenamefont {Novellino}, \citenamefont {Spada}, \citenamefont {Randazzo},\ and\ \citenamefont {da~Silva}}]{karsisiotis2011topological}%
  \BibitemOpen
  \bibfield  {author} {\bibinfo {author} {\bibfnamefont {A.~I.}\ \bibnamefont {Karsisiotis}}, \bibinfo {author} {\bibfnamefont {N.~M.}\ \bibnamefont {Hessari}}, \bibinfo {author} {\bibfnamefont {E.}~\bibnamefont {Novellino}}, \bibinfo {author} {\bibfnamefont {G.~P.}\ \bibnamefont {Spada}}, \bibinfo {author} {\bibfnamefont {A.}~\bibnamefont {Randazzo}},\ and\ \bibinfo {author} {\bibfnamefont {M.~W.}\ \bibnamefont {da~Silva}},\ }\bibfield  {title} {\bibinfo {title} {Topological characterization of nucleic acid g-quadruplexes by uv absorption and circular dichroism},\ }\href@noop {} {\bibfield  {journal} {\bibinfo  {journal} {Angewandte Chemie}\ }\textbf {\bibinfo {volume} {45}},\ \bibinfo {pages} {10833} (\bibinfo {year} {2011})}\BibitemShut {NoStop}%
\bibitem [{\citenamefont {Lim}\ \emph {et~al.}(2009)\citenamefont {Lim}, \citenamefont {Amrane}, \citenamefont {Bouaziz}, \citenamefont {Xu}, \citenamefont {Mu}, \citenamefont {Patel}, \citenamefont {Luu},\ and\ \citenamefont {Phan}}]{lim2009structure}%
  \BibitemOpen
  \bibfield  {author} {\bibinfo {author} {\bibfnamefont {K.~W.}\ \bibnamefont {Lim}}, \bibinfo {author} {\bibfnamefont {S.}~\bibnamefont {Amrane}}, \bibinfo {author} {\bibfnamefont {S.}~\bibnamefont {Bouaziz}}, \bibinfo {author} {\bibfnamefont {W.}~\bibnamefont {Xu}}, \bibinfo {author} {\bibfnamefont {Y.}~\bibnamefont {Mu}}, \bibinfo {author} {\bibfnamefont {D.~J.}\ \bibnamefont {Patel}}, \bibinfo {author} {\bibfnamefont {K.~N.}\ \bibnamefont {Luu}},\ and\ \bibinfo {author} {\bibfnamefont {A.~T.}\ \bibnamefont {Phan}},\ }\bibfield  {title} {\bibinfo {title} {Structure of the human telomere in k+ solution: a stable basket-type g-quadruplex with only two g-tetrad layers},\ }\href@noop {} {\bibfield  {journal} {\bibinfo  {journal} {Journal of the American Chemical Society}\ }\textbf {\bibinfo {volume} {131}},\ \bibinfo {pages} {4301} (\bibinfo {year} {2009})}\BibitemShut {NoStop}%
\bibitem [{\citenamefont {Lim}\ \emph {et~al.}(2013)\citenamefont {Lim}, \citenamefont {Ng}, \citenamefont {Mart{\'\i}n-Pintado}, \citenamefont {Heddi},\ and\ \citenamefont {Phan}}]{lim2013structure}%
  \BibitemOpen
  \bibfield  {author} {\bibinfo {author} {\bibfnamefont {K.~W.}\ \bibnamefont {Lim}}, \bibinfo {author} {\bibfnamefont {V.~C.~M.}\ \bibnamefont {Ng}}, \bibinfo {author} {\bibfnamefont {N.}~\bibnamefont {Mart{\'\i}n-Pintado}}, \bibinfo {author} {\bibfnamefont {B.}~\bibnamefont {Heddi}},\ and\ \bibinfo {author} {\bibfnamefont {A.~T.}\ \bibnamefont {Phan}},\ }\bibfield  {title} {\bibinfo {title} {Structure of the human telomere in na+ solution: an antiparallel (2+ 2) g-quadruplex scaffold reveals additional diversity},\ }\href@noop {} {\bibfield  {journal} {\bibinfo  {journal} {Nucleic acids research}\ }\textbf {\bibinfo {volume} {41}},\ \bibinfo {pages} {10556} (\bibinfo {year} {2013})}\BibitemShut {NoStop}%
\bibitem [{\citenamefont {Wang}\ and\ \citenamefont {Patel}(1993)}]{wang1993solution}%
  \BibitemOpen
  \bibfield  {author} {\bibinfo {author} {\bibfnamefont {Y.}~\bibnamefont {Wang}}\ and\ \bibinfo {author} {\bibfnamefont {D.~J.}\ \bibnamefont {Patel}},\ }\bibfield  {title} {\bibinfo {title} {Solution structure of the human telomeric repeat d [ag3 (t2ag3) 3] g-tetraplex},\ }\href@noop {} {\bibfield  {journal} {\bibinfo  {journal} {Structure}\ }\textbf {\bibinfo {volume} {1}},\ \bibinfo {pages} {263} (\bibinfo {year} {1993})}\BibitemShut {NoStop}%
\bibitem [{\citenamefont {Parkinson}\ \emph {et~al.}(2002)\citenamefont {Parkinson}, \citenamefont {Lee},\ and\ \citenamefont {Neidle}}]{parkinson2002crystal}%
  \BibitemOpen
  \bibfield  {author} {\bibinfo {author} {\bibfnamefont {G.~N.}\ \bibnamefont {Parkinson}}, \bibinfo {author} {\bibfnamefont {M.~P.}\ \bibnamefont {Lee}},\ and\ \bibinfo {author} {\bibfnamefont {S.}~\bibnamefont {Neidle}},\ }\bibfield  {title} {\bibinfo {title} {Crystal structure of parallel quadruplexes from human telomeric dna},\ }\href@noop {} {\bibfield  {journal} {\bibinfo  {journal} {Nature}\ }\textbf {\bibinfo {volume} {417}},\ \bibinfo {pages} {876} (\bibinfo {year} {2002})}\BibitemShut {NoStop}%
\bibitem [{\citenamefont {Ambrus}\ \emph {et~al.}(2006)\citenamefont {Ambrus}, \citenamefont {Chen}, \citenamefont {Dai}, \citenamefont {Bialis}, \citenamefont {Jones},\ and\ \citenamefont {Yang}}]{ambrus2006human}%
  \BibitemOpen
  \bibfield  {author} {\bibinfo {author} {\bibfnamefont {A.}~\bibnamefont {Ambrus}}, \bibinfo {author} {\bibfnamefont {D.}~\bibnamefont {Chen}}, \bibinfo {author} {\bibfnamefont {J.}~\bibnamefont {Dai}}, \bibinfo {author} {\bibfnamefont {T.}~\bibnamefont {Bialis}}, \bibinfo {author} {\bibfnamefont {R.~A.}\ \bibnamefont {Jones}},\ and\ \bibinfo {author} {\bibfnamefont {D.}~\bibnamefont {Yang}},\ }\bibfield  {title} {\bibinfo {title} {Human telomeric sequence forms a hybrid-type intramolecular g-quadruplex structure with mixed parallel/antiparallel strands in potassium solution},\ }\href@noop {} {\bibfield  {journal} {\bibinfo  {journal} {Nucleic acids research}\ }\textbf {\bibinfo {volume} {34}},\ \bibinfo {pages} {2723} (\bibinfo {year} {2006})}\BibitemShut {NoStop}%
\bibitem [{\citenamefont {Dai}\ \emph {et~al.}(2007{\natexlab{a}})\citenamefont {Dai}, \citenamefont {Carver}, \citenamefont {Punchihewa}, \citenamefont {Jones},\ and\ \citenamefont {Yang}}]{dai2007structure}%
  \BibitemOpen
  \bibfield  {author} {\bibinfo {author} {\bibfnamefont {J.}~\bibnamefont {Dai}}, \bibinfo {author} {\bibfnamefont {M.}~\bibnamefont {Carver}}, \bibinfo {author} {\bibfnamefont {C.}~\bibnamefont {Punchihewa}}, \bibinfo {author} {\bibfnamefont {R.~A.}\ \bibnamefont {Jones}},\ and\ \bibinfo {author} {\bibfnamefont {D.}~\bibnamefont {Yang}},\ }\bibfield  {title} {\bibinfo {title} {Structure of the hybrid-2 type intramolecular human telomeric g-quadruplex in k+ solution: insights into structure polymorphism of the human telomeric sequence},\ }\href@noop {} {\bibfield  {journal} {\bibinfo  {journal} {Nucleic acids research}\ }\textbf {\bibinfo {volume} {35}},\ \bibinfo {pages} {4927} (\bibinfo {year} {2007}{\natexlab{a}})}\BibitemShut {NoStop}%
\bibitem [{\citenamefont {Dai}\ \emph {et~al.}(2007{\natexlab{b}})\citenamefont {Dai}, \citenamefont {Punchihewa}, \citenamefont {Ambrus}, \citenamefont {Chen}, \citenamefont {Jones},\ and\ \citenamefont {Yang}}]{dai2007structure2}%
  \BibitemOpen
  \bibfield  {author} {\bibinfo {author} {\bibfnamefont {J.}~\bibnamefont {Dai}}, \bibinfo {author} {\bibfnamefont {C.}~\bibnamefont {Punchihewa}}, \bibinfo {author} {\bibfnamefont {A.}~\bibnamefont {Ambrus}}, \bibinfo {author} {\bibfnamefont {D.}~\bibnamefont {Chen}}, \bibinfo {author} {\bibfnamefont {R.~A.}\ \bibnamefont {Jones}},\ and\ \bibinfo {author} {\bibfnamefont {D.}~\bibnamefont {Yang}},\ }\bibfield  {title} {\bibinfo {title} {Structure of the intramolecular human telomeric g-quadruplex in potassium solution: a novel adenine triple formation},\ }\href@noop {} {\bibfield  {journal} {\bibinfo  {journal} {Nucleic acids research}\ }\textbf {\bibinfo {volume} {35}},\ \bibinfo {pages} {2440} (\bibinfo {year} {2007}{\natexlab{b}})}\BibitemShut {NoStop}%
\bibitem [{\citenamefont {Luu}\ \emph {et~al.}(2006)\citenamefont {Luu}, \citenamefont {Phan}, \citenamefont {Kuryavyi}, \citenamefont {Lacroix},\ and\ \citenamefont {Patel}}]{luu2006structure}%
  \BibitemOpen
  \bibfield  {author} {\bibinfo {author} {\bibfnamefont {K.~N.}\ \bibnamefont {Luu}}, \bibinfo {author} {\bibfnamefont {A.~T.}\ \bibnamefont {Phan}}, \bibinfo {author} {\bibfnamefont {V.}~\bibnamefont {Kuryavyi}}, \bibinfo {author} {\bibfnamefont {L.}~\bibnamefont {Lacroix}},\ and\ \bibinfo {author} {\bibfnamefont {D.~J.}\ \bibnamefont {Patel}},\ }\bibfield  {title} {\bibinfo {title} {Structure of the human telomere in k+ solution: an intramolecular (3+ 1) g-quadruplex scaffold},\ }\href@noop {} {\bibfield  {journal} {\bibinfo  {journal} {Journal of the American Chemical Society}\ }\textbf {\bibinfo {volume} {128}},\ \bibinfo {pages} {9963} (\bibinfo {year} {2006})}\BibitemShut {NoStop}%
\bibitem [{\citenamefont {Phan}\ \emph {et~al.}(2007)\citenamefont {Phan}, \citenamefont {Kuryavyi}, \citenamefont {Luu},\ and\ \citenamefont {Patel}}]{phan2007structure}%
  \BibitemOpen
  \bibfield  {author} {\bibinfo {author} {\bibfnamefont {A.~T.}\ \bibnamefont {Phan}}, \bibinfo {author} {\bibfnamefont {V.}~\bibnamefont {Kuryavyi}}, \bibinfo {author} {\bibfnamefont {K.~N.}\ \bibnamefont {Luu}},\ and\ \bibinfo {author} {\bibfnamefont {D.~J.}\ \bibnamefont {Patel}},\ }\bibfield  {title} {\bibinfo {title} {Structure of two intramolecular g-quadruplexes formed by natural human telomere sequences in k+ solution},\ }\href@noop {} {\bibfield  {journal} {\bibinfo  {journal} {Nucleic acids research}\ }\textbf {\bibinfo {volume} {35}},\ \bibinfo {pages} {6517} (\bibinfo {year} {2007})}\BibitemShut {NoStop}%
\bibitem [{\citenamefont {Zhang}\ \emph {et~al.}(2010)\citenamefont {Zhang}, \citenamefont {Dai}, \citenamefont {Veliath}, \citenamefont {Jones},\ and\ \citenamefont {Yang}}]{zhang2010structure}%
  \BibitemOpen
  \bibfield  {author} {\bibinfo {author} {\bibfnamefont {Z.}~\bibnamefont {Zhang}}, \bibinfo {author} {\bibfnamefont {J.}~\bibnamefont {Dai}}, \bibinfo {author} {\bibfnamefont {E.}~\bibnamefont {Veliath}}, \bibinfo {author} {\bibfnamefont {R.~A.}\ \bibnamefont {Jones}},\ and\ \bibinfo {author} {\bibfnamefont {D.}~\bibnamefont {Yang}},\ }\bibfield  {title} {\bibinfo {title} {Structure of a two-g-tetrad intramolecular g-quadruplex formed by a variant human telomeric sequence in k+ solution: insights into the interconversion of human telomeric g-quadruplex structures},\ }\href@noop {} {\bibfield  {journal} {\bibinfo  {journal} {Nucleic acids research}\ }\textbf {\bibinfo {volume} {38}},\ \bibinfo {pages} {1009} (\bibinfo {year} {2010})}\BibitemShut {NoStop}%
\bibitem [{\citenamefont {Phan}\ \emph {et~al.}(2006{\natexlab{a}})\citenamefont {Phan}, \citenamefont {Luu},\ and\ \citenamefont {Patel}}]{phan2006different}%
  \BibitemOpen
  \bibfield  {author} {\bibinfo {author} {\bibfnamefont {A.~T.}\ \bibnamefont {Phan}}, \bibinfo {author} {\bibfnamefont {K.~N.}\ \bibnamefont {Luu}},\ and\ \bibinfo {author} {\bibfnamefont {D.~J.}\ \bibnamefont {Patel}},\ }\bibfield  {title} {\bibinfo {title} {Different loop arrangements of intramolecular human telomeric (3+ 1) g-quadruplexes in k+ solution},\ }\href@noop {} {\bibfield  {journal} {\bibinfo  {journal} {Nucleic acids research}\ }\textbf {\bibinfo {volume} {34}},\ \bibinfo {pages} {5715} (\bibinfo {year} {2006}{\natexlab{a}})}\BibitemShut {NoStop}%
\bibitem [{\citenamefont {Li}\ \emph {et~al.}(2005)\citenamefont {Li}, \citenamefont {Correia}, \citenamefont {Wang}, \citenamefont {Trent},\ and\ \citenamefont {Chaires}}]{li2005not}%
  \BibitemOpen
  \bibfield  {author} {\bibinfo {author} {\bibfnamefont {J.}~\bibnamefont {Li}}, \bibinfo {author} {\bibfnamefont {J.~J.}\ \bibnamefont {Correia}}, \bibinfo {author} {\bibfnamefont {L.}~\bibnamefont {Wang}}, \bibinfo {author} {\bibfnamefont {J.~O.}\ \bibnamefont {Trent}},\ and\ \bibinfo {author} {\bibfnamefont {J.~B.}\ \bibnamefont {Chaires}},\ }\bibfield  {title} {\bibinfo {title} {Not so crystal clear: the structure of the human telomere g-quadruplex in solution differs from that present in a crystal},\ }\href@noop {} {\bibfield  {journal} {\bibinfo  {journal} {Nucleic acids research}\ }\textbf {\bibinfo {volume} {33}},\ \bibinfo {pages} {4649} (\bibinfo {year} {2005})}\BibitemShut {NoStop}%
\bibitem [{\citenamefont {Dai}\ \emph {et~al.}(2008)\citenamefont {Dai}, \citenamefont {Carver},\ and\ \citenamefont {Yang}}]{dai2008polymorphism}%
  \BibitemOpen
  \bibfield  {author} {\bibinfo {author} {\bibfnamefont {J.}~\bibnamefont {Dai}}, \bibinfo {author} {\bibfnamefont {M.}~\bibnamefont {Carver}},\ and\ \bibinfo {author} {\bibfnamefont {D.}~\bibnamefont {Yang}},\ }\bibfield  {title} {\bibinfo {title} {Polymorphism of human telomeric quadruplex structures},\ }\href@noop {} {\bibfield  {journal} {\bibinfo  {journal} {Biochimie}\ }\textbf {\bibinfo {volume} {90}},\ \bibinfo {pages} {1172} (\bibinfo {year} {2008})}\BibitemShut {NoStop}%
\bibitem [{\citenamefont {Long}\ and\ \citenamefont {Stone}(2013)}]{long2013kinetic}%
  \BibitemOpen
  \bibfield  {author} {\bibinfo {author} {\bibfnamefont {X.}~\bibnamefont {Long}}\ and\ \bibinfo {author} {\bibfnamefont {M.~D.}\ \bibnamefont {Stone}},\ }\bibfield  {title} {\bibinfo {title} {Kinetic partitioning modulates human telomere dna g-quadruplex structural polymorphism},\ }\href@noop {} {\bibfield  {journal} {\bibinfo  {journal} {PLoS One}\ }\textbf {\bibinfo {volume} {8}},\ \bibinfo {pages} {e83420} (\bibinfo {year} {2013})}\BibitemShut {NoStop}%
\bibitem [{\citenamefont {Gray}\ \emph {et~al.}(2014)\citenamefont {Gray}, \citenamefont {Trent},\ and\ \citenamefont {Chaires}}]{gray2014folding}%
  \BibitemOpen
  \bibfield  {author} {\bibinfo {author} {\bibfnamefont {R.~D.}\ \bibnamefont {Gray}}, \bibinfo {author} {\bibfnamefont {J.~O.}\ \bibnamefont {Trent}},\ and\ \bibinfo {author} {\bibfnamefont {J.~B.}\ \bibnamefont {Chaires}},\ }\bibfield  {title} {\bibinfo {title} {Folding and unfolding pathways of the human telomeric g-quadruplex},\ }\href@noop {} {\bibfield  {journal} {\bibinfo  {journal} {Journal of molecular biology}\ }\textbf {\bibinfo {volume} {426}},\ \bibinfo {pages} {1629} (\bibinfo {year} {2014})}\BibitemShut {NoStop}%
\bibitem [{\citenamefont {You}\ \emph {et~al.}(2014)\citenamefont {You}, \citenamefont {Zeng}, \citenamefont {Xu}, \citenamefont {Lim}, \citenamefont {Efremov}, \citenamefont {Phan},\ and\ \citenamefont {Yan}}]{you2014dynamics}%
  \BibitemOpen
  \bibfield  {author} {\bibinfo {author} {\bibfnamefont {H.}~\bibnamefont {You}}, \bibinfo {author} {\bibfnamefont {X.}~\bibnamefont {Zeng}}, \bibinfo {author} {\bibfnamefont {Y.}~\bibnamefont {Xu}}, \bibinfo {author} {\bibfnamefont {C.~J.}\ \bibnamefont {Lim}}, \bibinfo {author} {\bibfnamefont {A.~K.}\ \bibnamefont {Efremov}}, \bibinfo {author} {\bibfnamefont {A.~T.}\ \bibnamefont {Phan}},\ and\ \bibinfo {author} {\bibfnamefont {J.}~\bibnamefont {Yan}},\ }\bibfield  {title} {\bibinfo {title} {Dynamics and stability of polymorphic human telomeric g-quadruplex under tension},\ }\href@noop {} {\bibfield  {journal} {\bibinfo  {journal} {Nucleic acids research}\ }\textbf {\bibinfo {volume} {42}},\ \bibinfo {pages} {8789} (\bibinfo {year} {2014})}\BibitemShut {NoStop}%
\bibitem [{\citenamefont {Bessi}\ \emph {et~al.}(2015)\citenamefont {Bessi}, \citenamefont {Jonker}, \citenamefont {Richter},\ and\ \citenamefont {Schwalbe}}]{bessi2015involvement}%
  \BibitemOpen
  \bibfield  {author} {\bibinfo {author} {\bibfnamefont {I.}~\bibnamefont {Bessi}}, \bibinfo {author} {\bibfnamefont {H.~R.}\ \bibnamefont {Jonker}}, \bibinfo {author} {\bibfnamefont {C.}~\bibnamefont {Richter}},\ and\ \bibinfo {author} {\bibfnamefont {H.}~\bibnamefont {Schwalbe}},\ }\bibfield  {title} {\bibinfo {title} {Involvement of long-lived intermediate states in the complex folding pathway of the human telomeric g-quadruplex},\ }\href@noop {} {\bibfield  {journal} {\bibinfo  {journal} {Angewandte Chemie}\ }\textbf {\bibinfo {volume} {127}},\ \bibinfo {pages} {8564} (\bibinfo {year} {2015})}\BibitemShut {NoStop}%
\bibitem [{\citenamefont {Armstrong}\ \emph {et~al.}(2015)\citenamefont {Armstrong}, \citenamefont {Riskowski},\ and\ \citenamefont {Strouse}}]{armstrong2015nanometal}%
  \BibitemOpen
  \bibfield  {author} {\bibinfo {author} {\bibfnamefont {R.~E.}\ \bibnamefont {Armstrong}}, \bibinfo {author} {\bibfnamefont {R.~A.}\ \bibnamefont {Riskowski}},\ and\ \bibinfo {author} {\bibfnamefont {G.~F.}\ \bibnamefont {Strouse}},\ }\bibfield  {title} {\bibinfo {title} {Nanometal surface energy transfer optical ruler for measuring a human telomere structure},\ }\href@noop {} {\bibfield  {journal} {\bibinfo  {journal} {Photochemistry and photobiology}\ }\textbf {\bibinfo {volume} {91}},\ \bibinfo {pages} {732} (\bibinfo {year} {2015})}\BibitemShut {NoStop}%
\bibitem [{\citenamefont {Xue}\ \emph {et~al.}(2011)\citenamefont {Xue}, \citenamefont {Liu}, \citenamefont {Zheng}, \citenamefont {Kan}, \citenamefont {Hao},\ and\ \citenamefont {Tan}}]{xue2011kinetic}%
  \BibitemOpen
  \bibfield  {author} {\bibinfo {author} {\bibfnamefont {Y.}~\bibnamefont {Xue}}, \bibinfo {author} {\bibfnamefont {J.-q.}\ \bibnamefont {Liu}}, \bibinfo {author} {\bibfnamefont {K.-w.}\ \bibnamefont {Zheng}}, \bibinfo {author} {\bibfnamefont {Z.-y.}\ \bibnamefont {Kan}}, \bibinfo {author} {\bibfnamefont {Y.-h.}\ \bibnamefont {Hao}},\ and\ \bibinfo {author} {\bibfnamefont {Z.}~\bibnamefont {Tan}},\ }\bibfield  {title} {\bibinfo {title} {Kinetic and thermodynamic control of g-quadruplex folding},\ }\href@noop {} {\bibfield  {journal} {\bibinfo  {journal} {Angewandte Chemie International Edition}\ }\textbf {\bibinfo {volume} {50}},\ \bibinfo {pages} {8046} (\bibinfo {year} {2011})}\BibitemShut {NoStop}%
\bibitem [{\citenamefont {Lannan}\ \emph {et~al.}(2012)\citenamefont {Lannan}, \citenamefont {Mamajanov},\ and\ \citenamefont {Hud}}]{lannan2012human}%
  \BibitemOpen
  \bibfield  {author} {\bibinfo {author} {\bibfnamefont {F.~M.}\ \bibnamefont {Lannan}}, \bibinfo {author} {\bibfnamefont {I.}~\bibnamefont {Mamajanov}},\ and\ \bibinfo {author} {\bibfnamefont {N.~V.}\ \bibnamefont {Hud}},\ }\bibfield  {title} {\bibinfo {title} {Human telomere sequence dna in water-free and high-viscosity solvents: G-quadruplex folding governed by kramers rate theory},\ }\href@noop {} {\bibfield  {journal} {\bibinfo  {journal} {Journal of the American Chemical Society}\ }\textbf {\bibinfo {volume} {134}},\ \bibinfo {pages} {15324} (\bibinfo {year} {2012})}\BibitemShut {NoStop}%
\bibitem [{\citenamefont {Gabelica}(2014)}]{gabelica2014pilgrim}%
  \BibitemOpen
  \bibfield  {author} {\bibinfo {author} {\bibfnamefont {V.}~\bibnamefont {Gabelica}},\ }\bibfield  {title} {\bibinfo {title} {A pilgrim's guide to g-quadruplex nucleic acid folding},\ }\href@noop {} {\bibfield  {journal} {\bibinfo  {journal} {Biochimie}\ }\textbf {\bibinfo {volume} {105}},\ \bibinfo {pages} {1} (\bibinfo {year} {2014})}\BibitemShut {NoStop}%
\bibitem [{\citenamefont {Lane}\ \emph {et~al.}(2008)\citenamefont {Lane}, \citenamefont {Chaires}, \citenamefont {Gray},\ and\ \citenamefont {Trent}}]{lane2008stability}%
  \BibitemOpen
  \bibfield  {author} {\bibinfo {author} {\bibfnamefont {A.~N.}\ \bibnamefont {Lane}}, \bibinfo {author} {\bibfnamefont {J.~B.}\ \bibnamefont {Chaires}}, \bibinfo {author} {\bibfnamefont {R.~D.}\ \bibnamefont {Gray}},\ and\ \bibinfo {author} {\bibfnamefont {J.~O.}\ \bibnamefont {Trent}},\ }\bibfield  {title} {\bibinfo {title} {Stability and kinetics of g-quadruplex structures},\ }\href@noop {} {\bibfield  {journal} {\bibinfo  {journal} {Nucleic acids research}\ }\textbf {\bibinfo {volume} {36}},\ \bibinfo {pages} {5482} (\bibinfo {year} {2008})}\BibitemShut {NoStop}%
\bibitem [{\citenamefont {Biffi}\ \emph {et~al.}(2013)\citenamefont {Biffi}, \citenamefont {Tannahill}, \citenamefont {McCafferty},\ and\ \citenamefont {Balasubramanian}}]{biffi2013quantitative}%
  \BibitemOpen
  \bibfield  {author} {\bibinfo {author} {\bibfnamefont {G.}~\bibnamefont {Biffi}}, \bibinfo {author} {\bibfnamefont {D.}~\bibnamefont {Tannahill}}, \bibinfo {author} {\bibfnamefont {J.}~\bibnamefont {McCafferty}},\ and\ \bibinfo {author} {\bibfnamefont {S.}~\bibnamefont {Balasubramanian}},\ }\bibfield  {title} {\bibinfo {title} {Quantitative visualization of dna g-quadruplex structures in human cells},\ }\href@noop {} {\bibfield  {journal} {\bibinfo  {journal} {Nature chemistry}\ }\textbf {\bibinfo {volume} {5}},\ \bibinfo {pages} {182} (\bibinfo {year} {2013})}\BibitemShut {NoStop}%
\bibitem [{\citenamefont {Huppert}(2010)}]{huppert2010structure}%
  \BibitemOpen
  \bibfield  {author} {\bibinfo {author} {\bibfnamefont {J.~L.}\ \bibnamefont {Huppert}},\ }\bibfield  {title} {\bibinfo {title} {Structure, location and interactions of g-quadruplexes},\ }\href@noop {} {\bibfield  {journal} {\bibinfo  {journal} {The FEBS journal}\ }\textbf {\bibinfo {volume} {277}},\ \bibinfo {pages} {3452} (\bibinfo {year} {2010})}\BibitemShut {NoStop}%
\bibitem [{\citenamefont {Phan}\ \emph {et~al.}(2006{\natexlab{b}})\citenamefont {Phan}, \citenamefont {Kuryavyi},\ and\ \citenamefont {Patel}}]{phan2006dna}%
  \BibitemOpen
  \bibfield  {author} {\bibinfo {author} {\bibfnamefont {A.~T.}\ \bibnamefont {Phan}}, \bibinfo {author} {\bibfnamefont {V.}~\bibnamefont {Kuryavyi}},\ and\ \bibinfo {author} {\bibfnamefont {D.~J.}\ \bibnamefont {Patel}},\ }\bibfield  {title} {\bibinfo {title} {Dna architecture: from g to z},\ }\href@noop {} {\bibfield  {journal} {\bibinfo  {journal} {Current opinion in structural biology}\ }\textbf {\bibinfo {volume} {16}},\ \bibinfo {pages} {288} (\bibinfo {year} {2006}{\natexlab{b}})}\BibitemShut {NoStop}%
\bibitem [{\citenamefont {V{\'\i}glask{\`y}}\ \emph {et~al.}(2011)\citenamefont {V{\'\i}glask{\`y}}, \citenamefont {Tlu{\v{c}}kov{\'a}},\ and\ \citenamefont {Bauer}}]{viglasky2011first}%
  \BibitemOpen
  \bibfield  {author} {\bibinfo {author} {\bibfnamefont {V.}~\bibnamefont {V{\'\i}glask{\`y}}}, \bibinfo {author} {\bibfnamefont {K.}~\bibnamefont {Tlu{\v{c}}kov{\'a}}},\ and\ \bibinfo {author} {\bibfnamefont {L.}~\bibnamefont {Bauer}},\ }\bibfield  {title} {\bibinfo {title} {The first derivative of a function of circular dichroism spectra: biophysical study of human telomeric g-quadruplex},\ }\href@noop {} {\bibfield  {journal} {\bibinfo  {journal} {European Biophysics Journal}\ }\textbf {\bibinfo {volume} {40}},\ \bibinfo {pages} {29} (\bibinfo {year} {2011})}\BibitemShut {NoStop}%
\bibitem [{\citenamefont {Chaires}(2010)}]{chaires2010human}%
  \BibitemOpen
  \bibfield  {author} {\bibinfo {author} {\bibfnamefont {J.~B.}\ \bibnamefont {Chaires}},\ }\bibfield  {title} {\bibinfo {title} {Human telomeric g-quadruplex: thermodynamic and kinetic studies of telomeric quadruplex stability},\ }\href@noop {} {\bibfield  {journal} {\bibinfo  {journal} {The FEBS journal}\ }\textbf {\bibinfo {volume} {277}},\ \bibinfo {pages} {1098} (\bibinfo {year} {2010})}\BibitemShut {NoStop}%
\bibitem [{\citenamefont {Smargiasso}\ \emph {et~al.}(2008)\citenamefont {Smargiasso}, \citenamefont {Rosu}, \citenamefont {Hsia}, \citenamefont {Colson}, \citenamefont {Baker}, \citenamefont {Bowers}, \citenamefont {De~Pauw},\ and\ \citenamefont {Gabelica}}]{smargiasso2008g}%
  \BibitemOpen
  \bibfield  {author} {\bibinfo {author} {\bibfnamefont {N.}~\bibnamefont {Smargiasso}}, \bibinfo {author} {\bibfnamefont {F.}~\bibnamefont {Rosu}}, \bibinfo {author} {\bibfnamefont {W.}~\bibnamefont {Hsia}}, \bibinfo {author} {\bibfnamefont {P.}~\bibnamefont {Colson}}, \bibinfo {author} {\bibfnamefont {E.~S.}\ \bibnamefont {Baker}}, \bibinfo {author} {\bibfnamefont {M.~T.}\ \bibnamefont {Bowers}}, \bibinfo {author} {\bibfnamefont {E.}~\bibnamefont {De~Pauw}},\ and\ \bibinfo {author} {\bibfnamefont {V.}~\bibnamefont {Gabelica}},\ }\bibfield  {title} {\bibinfo {title} {G-quadruplex dna assemblies: loop length, cation identity, and multimer formation},\ }\href@noop {} {\bibfield  {journal} {\bibinfo  {journal} {Journal of the American Chemical Society}\ }\textbf {\bibinfo {volume} {130}},\ \bibinfo {pages} {10208} (\bibinfo {year} {2008})}\BibitemShut {NoStop}%
\bibitem [{\citenamefont {Heddi}\ and\ \citenamefont {Phan}(2011)}]{heddi2011structure}%
  \BibitemOpen
  \bibfield  {author} {\bibinfo {author} {\bibfnamefont {B.}~\bibnamefont {Heddi}}\ and\ \bibinfo {author} {\bibfnamefont {A.~T.}\ \bibnamefont {Phan}},\ }\bibfield  {title} {\bibinfo {title} {Structure of human telomeric dna in crowded solution},\ }\href@noop {} {\bibfield  {journal} {\bibinfo  {journal} {Journal of the American Chemical Society}\ }\textbf {\bibinfo {volume} {133}},\ \bibinfo {pages} {9824} (\bibinfo {year} {2011})}\BibitemShut {NoStop}%
\bibitem [{\citenamefont {Huppert}\ and\ \citenamefont {Balasubramanian}(2005)}]{huppert2005prevalence}%
  \BibitemOpen
  \bibfield  {author} {\bibinfo {author} {\bibfnamefont {J.~L.}\ \bibnamefont {Huppert}}\ and\ \bibinfo {author} {\bibfnamefont {S.}~\bibnamefont {Balasubramanian}},\ }\bibfield  {title} {\bibinfo {title} {Prevalence of quadruplexes in the human genome},\ }\href@noop {} {\bibfield  {journal} {\bibinfo  {journal} {Nucleic acids research}\ }\textbf {\bibinfo {volume} {33}},\ \bibinfo {pages} {2908} (\bibinfo {year} {2005})}\BibitemShut {NoStop}%
\bibitem [{\citenamefont {Todd}\ \emph {et~al.}(2005)\citenamefont {Todd}, \citenamefont {Johnston},\ and\ \citenamefont {Neidle}}]{todd2005highly}%
  \BibitemOpen
  \bibfield  {author} {\bibinfo {author} {\bibfnamefont {A.~K.}\ \bibnamefont {Todd}}, \bibinfo {author} {\bibfnamefont {M.}~\bibnamefont {Johnston}},\ and\ \bibinfo {author} {\bibfnamefont {S.}~\bibnamefont {Neidle}},\ }\bibfield  {title} {\bibinfo {title} {Highly prevalent putative quadruplex sequence motifs in human dna},\ }\href@noop {} {\bibfield  {journal} {\bibinfo  {journal} {Nucleic acids research}\ }\textbf {\bibinfo {volume} {33}},\ \bibinfo {pages} {2901} (\bibinfo {year} {2005})}\BibitemShut {NoStop}%
\bibitem [{\citenamefont {H{\"a}nsel-Hertsch}\ \emph {et~al.}(2017)\citenamefont {H{\"a}nsel-Hertsch}, \citenamefont {Di~Antonio},\ and\ \citenamefont {Balasubramanian}}]{hansel2017dna}%
  \BibitemOpen
  \bibfield  {author} {\bibinfo {author} {\bibfnamefont {R.}~\bibnamefont {H{\"a}nsel-Hertsch}}, \bibinfo {author} {\bibfnamefont {M.}~\bibnamefont {Di~Antonio}},\ and\ \bibinfo {author} {\bibfnamefont {S.}~\bibnamefont {Balasubramanian}},\ }\bibfield  {title} {\bibinfo {title} {Dna g-quadruplexes in the human genome: detection, functions and therapeutic potential},\ }\href@noop {} {\bibfield  {journal} {\bibinfo  {journal} {Nature reviews Molecular cell biology}\ }\textbf {\bibinfo {volume} {18}},\ \bibinfo {pages} {279} (\bibinfo {year} {2017})}\BibitemShut {NoStop}%
\bibitem [{\citenamefont {Schaffitzel}\ \emph {et~al.}(2001)\citenamefont {Schaffitzel}, \citenamefont {Berger}, \citenamefont {Postberg}, \citenamefont {Hanes}, \citenamefont {Lipps},\ and\ \citenamefont {Pl{\"u}ckthun}}]{schaffitzel2001vitro}%
  \BibitemOpen
  \bibfield  {author} {\bibinfo {author} {\bibfnamefont {C.}~\bibnamefont {Schaffitzel}}, \bibinfo {author} {\bibfnamefont {I.}~\bibnamefont {Berger}}, \bibinfo {author} {\bibfnamefont {J.}~\bibnamefont {Postberg}}, \bibinfo {author} {\bibfnamefont {J.}~\bibnamefont {Hanes}}, \bibinfo {author} {\bibfnamefont {H.~J.}\ \bibnamefont {Lipps}},\ and\ \bibinfo {author} {\bibfnamefont {A.}~\bibnamefont {Pl{\"u}ckthun}},\ }\bibfield  {title} {\bibinfo {title} {In vitro generated antibodies specific for telomeric guanine-quadruplex dna react with stylonychia lemnae macronuclei},\ }\href@noop {} {\bibfield  {journal} {\bibinfo  {journal} {Proceedings of the National Academy of Sciences}\ }\textbf {\bibinfo {volume} {98}},\ \bibinfo {pages} {8572} (\bibinfo {year} {2001})}\BibitemShut {NoStop}%
\bibitem [{\citenamefont {Henderson}\ \emph {et~al.}(2013)\citenamefont {Henderson}, \citenamefont {Wu}, \citenamefont {Huang}, \citenamefont {Chavez}, \citenamefont {Platt}, \citenamefont {Johnson}, \citenamefont {Brosh}, \citenamefont {Sen},\ and\ \citenamefont {Lansdorp}}]{henderson2013detection}%
  \BibitemOpen
  \bibfield  {author} {\bibinfo {author} {\bibfnamefont {A.}~\bibnamefont {Henderson}}, \bibinfo {author} {\bibfnamefont {Y.}~\bibnamefont {Wu}}, \bibinfo {author} {\bibfnamefont {Y.~C.}\ \bibnamefont {Huang}}, \bibinfo {author} {\bibfnamefont {E.~A.}\ \bibnamefont {Chavez}}, \bibinfo {author} {\bibfnamefont {J.}~\bibnamefont {Platt}}, \bibinfo {author} {\bibfnamefont {F.~B.}\ \bibnamefont {Johnson}}, \bibinfo {author} {\bibfnamefont {R.~M.}\ \bibnamefont {Brosh}}, \bibinfo {author} {\bibfnamefont {D.}~\bibnamefont {Sen}},\ and\ \bibinfo {author} {\bibfnamefont {P.~M.}\ \bibnamefont {Lansdorp}},\ }\bibfield  {title} {\bibinfo {title} {Detection of g-quadruplex dna in mammalian cells},\ }\href@noop {} {\bibfield  {journal} {\bibinfo  {journal} {Nucleic acids research}\ }\textbf {\bibinfo {volume} {42}},\ \bibinfo {pages} {860} (\bibinfo {year} {2013})}\BibitemShut {NoStop}%
\bibitem [{\citenamefont {Bao}\ \emph {et~al.}(2019)\citenamefont {Bao}, \citenamefont {Liu},\ and\ \citenamefont {Xu}}]{bao2019hybrid}%
  \BibitemOpen
  \bibfield  {author} {\bibinfo {author} {\bibfnamefont {H.-L.}\ \bibnamefont {Bao}}, \bibinfo {author} {\bibfnamefont {H.-s.}\ \bibnamefont {Liu}},\ and\ \bibinfo {author} {\bibfnamefont {Y.}~\bibnamefont {Xu}},\ }\bibfield  {title} {\bibinfo {title} {Hybrid-type and two-tetrad antiparallel telomere dna g-quadruplex structures in living human cells},\ }\href@noop {} {\bibfield  {journal} {\bibinfo  {journal} {Nucleic acids research}\ }\textbf {\bibinfo {volume} {47}},\ \bibinfo {pages} {4940} (\bibinfo {year} {2019})}\BibitemShut {NoStop}%
\bibitem [{\citenamefont {Lam}\ \emph {et~al.}(2013)\citenamefont {Lam}, \citenamefont {Beraldi}, \citenamefont {Tannahill},\ and\ \citenamefont {Balasubramanian}}]{lam2013g}%
  \BibitemOpen
  \bibfield  {author} {\bibinfo {author} {\bibfnamefont {E.~Y.~N.}\ \bibnamefont {Lam}}, \bibinfo {author} {\bibfnamefont {D.}~\bibnamefont {Beraldi}}, \bibinfo {author} {\bibfnamefont {D.}~\bibnamefont {Tannahill}},\ and\ \bibinfo {author} {\bibfnamefont {S.}~\bibnamefont {Balasubramanian}},\ }\bibfield  {title} {\bibinfo {title} {G-quadruplex structures are stable and detectable in human genomic dna},\ }\href@noop {} {\bibfield  {journal} {\bibinfo  {journal} {Nature communications}\ }\textbf {\bibinfo {volume} {4}},\ \bibinfo {pages} {1796} (\bibinfo {year} {2013})}\BibitemShut {NoStop}%
\bibitem [{\citenamefont {Huppert}\ and\ \citenamefont {Balasubramanian}(2007)}]{huppert2007g}%
  \BibitemOpen
  \bibfield  {author} {\bibinfo {author} {\bibfnamefont {J.~L.}\ \bibnamefont {Huppert}}\ and\ \bibinfo {author} {\bibfnamefont {S.}~\bibnamefont {Balasubramanian}},\ }\bibfield  {title} {\bibinfo {title} {G-quadruplexes in promoters throughout the human genome},\ }\href@noop {} {\bibfield  {journal} {\bibinfo  {journal} {Nucleic acids research}\ }\textbf {\bibinfo {volume} {35}},\ \bibinfo {pages} {406} (\bibinfo {year} {2007})}\BibitemShut {NoStop}%
\bibitem [{\citenamefont {Kendrick}\ and\ \citenamefont {Hurley}(2010)}]{kendrick2010role}%
  \BibitemOpen
  \bibfield  {author} {\bibinfo {author} {\bibfnamefont {S.}~\bibnamefont {Kendrick}}\ and\ \bibinfo {author} {\bibfnamefont {L.~H.}\ \bibnamefont {Hurley}},\ }\bibfield  {title} {\bibinfo {title} {The role of g-quadruplex/i-motif secondary structures as cis-acting regulatory elements},\ }\href@noop {} {\bibfield  {journal} {\bibinfo  {journal} {Pure and Applied Chemistry}\ }\textbf {\bibinfo {volume} {82}},\ \bibinfo {pages} {1609} (\bibinfo {year} {2010})}\BibitemShut {NoStop}%
\bibitem [{\citenamefont {Rhodes}\ and\ \citenamefont {Lipps}(2015)}]{rhodes2015g}%
  \BibitemOpen
  \bibfield  {author} {\bibinfo {author} {\bibfnamefont {D.}~\bibnamefont {Rhodes}}\ and\ \bibinfo {author} {\bibfnamefont {H.~J.}\ \bibnamefont {Lipps}},\ }\bibfield  {title} {\bibinfo {title} {G-quadruplexes and their regulatory roles in biology},\ }\href@noop {} {\bibfield  {journal} {\bibinfo  {journal} {Nucleic acids research}\ }\textbf {\bibinfo {volume} {43}},\ \bibinfo {pages} {8627} (\bibinfo {year} {2015})}\BibitemShut {NoStop}%
\bibitem [{\citenamefont {Bianchi}\ \emph {et~al.}(2018)\citenamefont {Bianchi}, \citenamefont {Comez}, \citenamefont {Biehl}, \citenamefont {D’Amico}, \citenamefont {Gessini}, \citenamefont {Longo}, \citenamefont {Masciovecchio}, \citenamefont {Petrillo}, \citenamefont {Radulescu}, \citenamefont {Rossi} \emph {et~al.}}]{bianchi2018structure}%
  \BibitemOpen
  \bibfield  {author} {\bibinfo {author} {\bibfnamefont {F.}~\bibnamefont {Bianchi}}, \bibinfo {author} {\bibfnamefont {L.}~\bibnamefont {Comez}}, \bibinfo {author} {\bibfnamefont {R.}~\bibnamefont {Biehl}}, \bibinfo {author} {\bibfnamefont {F.}~\bibnamefont {D’Amico}}, \bibinfo {author} {\bibfnamefont {A.}~\bibnamefont {Gessini}}, \bibinfo {author} {\bibfnamefont {M.}~\bibnamefont {Longo}}, \bibinfo {author} {\bibfnamefont {C.}~\bibnamefont {Masciovecchio}}, \bibinfo {author} {\bibfnamefont {C.}~\bibnamefont {Petrillo}}, \bibinfo {author} {\bibfnamefont {A.}~\bibnamefont {Radulescu}}, \bibinfo {author} {\bibfnamefont {B.}~\bibnamefont {Rossi}}, \emph {et~al.},\ }\bibfield  {title} {\bibinfo {title} {Structure of human telomere g-quadruplex in the presence of a model drug along the thermal unfolding pathway},\ }\href@noop {} {\bibfield  {journal} {\bibinfo  {journal} {Nucleic acids research}\ }\textbf {\bibinfo {volume} {46}},\ \bibinfo {pages} {11927} (\bibinfo {year} {2018})}\BibitemShut {NoStop}%
\bibitem [{\citenamefont {Comez}\ \emph {et~al.}(2020)\citenamefont {Comez}, \citenamefont {Bianchi}, \citenamefont {Libera}, \citenamefont {Longo}, \citenamefont {Petrillo}, \citenamefont {Sacchetti}, \citenamefont {Sebastiani}, \citenamefont {D’Amico}, \citenamefont {Rossi}, \citenamefont {Gessini} \emph {et~al.}}]{comez2020polymorphism}%
  \BibitemOpen
  \bibfield  {author} {\bibinfo {author} {\bibfnamefont {L.}~\bibnamefont {Comez}}, \bibinfo {author} {\bibfnamefont {F.}~\bibnamefont {Bianchi}}, \bibinfo {author} {\bibfnamefont {V.}~\bibnamefont {Libera}}, \bibinfo {author} {\bibfnamefont {M.}~\bibnamefont {Longo}}, \bibinfo {author} {\bibfnamefont {C.}~\bibnamefont {Petrillo}}, \bibinfo {author} {\bibfnamefont {F.}~\bibnamefont {Sacchetti}}, \bibinfo {author} {\bibfnamefont {F.}~\bibnamefont {Sebastiani}}, \bibinfo {author} {\bibfnamefont {F.}~\bibnamefont {D’Amico}}, \bibinfo {author} {\bibfnamefont {B.}~\bibnamefont {Rossi}}, \bibinfo {author} {\bibfnamefont {A.}~\bibnamefont {Gessini}}, \emph {et~al.},\ }\bibfield  {title} {\bibinfo {title} {Polymorphism of human telomeric quadruplexes with drugs: A multi-technique biophysical study},\ }\href@noop {} {\bibfield  {journal} {\bibinfo  {journal} {Physical Chemistry Chemical Physics}\ }\textbf {\bibinfo {volume} {22}},\ \bibinfo {pages} {11583} (\bibinfo {year} {2020})}\BibitemShut {NoStop}%
\bibitem [{\citenamefont {Zahler}\ \emph {et~al.}(1991)\citenamefont {Zahler}, \citenamefont {Williamson}, \citenamefont {Cech},\ and\ \citenamefont {Prescott}}]{zahler1991inhibition}%
  \BibitemOpen
  \bibfield  {author} {\bibinfo {author} {\bibfnamefont {A.~M.}\ \bibnamefont {Zahler}}, \bibinfo {author} {\bibfnamefont {J.~R.}\ \bibnamefont {Williamson}}, \bibinfo {author} {\bibfnamefont {T.~R.}\ \bibnamefont {Cech}},\ and\ \bibinfo {author} {\bibfnamefont {D.~M.}\ \bibnamefont {Prescott}},\ }\bibfield  {title} {\bibinfo {title} {Inhibition of telomerase by g-quartet dma structures},\ }\href@noop {} {\bibfield  {journal} {\bibinfo  {journal} {Nature}\ }\textbf {\bibinfo {volume} {350}},\ \bibinfo {pages} {718} (\bibinfo {year} {1991})}\BibitemShut {NoStop}%
\bibitem [{\citenamefont {Tahara}\ \emph {et~al.}(2006)\citenamefont {Tahara}, \citenamefont {Shin-Ya}, \citenamefont {Seimiya}, \citenamefont {Yamada}, \citenamefont {Tsuruo},\ and\ \citenamefont {Ide}}]{tahara2006g}%
  \BibitemOpen
  \bibfield  {author} {\bibinfo {author} {\bibfnamefont {H.}~\bibnamefont {Tahara}}, \bibinfo {author} {\bibfnamefont {K.}~\bibnamefont {Shin-Ya}}, \bibinfo {author} {\bibfnamefont {H.}~\bibnamefont {Seimiya}}, \bibinfo {author} {\bibfnamefont {H.}~\bibnamefont {Yamada}}, \bibinfo {author} {\bibfnamefont {T.}~\bibnamefont {Tsuruo}},\ and\ \bibinfo {author} {\bibfnamefont {T.}~\bibnamefont {Ide}},\ }\bibfield  {title} {\bibinfo {title} {G-quadruplex stabilization by telomestatin induces trf2 protein dissociation from telomeres and anaphase bridge formation accompanied by loss of the 3 telomeric overhang in cancer cells},\ }\href@noop {} {\bibfield  {journal} {\bibinfo  {journal} {Oncogene}\ }\textbf {\bibinfo {volume} {25}},\ \bibinfo {pages} {1955} (\bibinfo {year} {2006})}\BibitemShut {NoStop}%
\bibitem [{\citenamefont {Zhou}\ \emph {et~al.}(2016)\citenamefont {Zhou}, \citenamefont {Liu}, \citenamefont {Li}, \citenamefont {Xu}, \citenamefont {Ma}, \citenamefont {Wu}, \citenamefont {Cheng}, \citenamefont {Yu}, \citenamefont {Zhao},\ and\ \citenamefont {Chen}}]{zhou2016telomere}%
  \BibitemOpen
  \bibfield  {author} {\bibinfo {author} {\bibfnamefont {G.}~\bibnamefont {Zhou}}, \bibinfo {author} {\bibfnamefont {X.}~\bibnamefont {Liu}}, \bibinfo {author} {\bibfnamefont {Y.}~\bibnamefont {Li}}, \bibinfo {author} {\bibfnamefont {S.}~\bibnamefont {Xu}}, \bibinfo {author} {\bibfnamefont {C.}~\bibnamefont {Ma}}, \bibinfo {author} {\bibfnamefont {X.}~\bibnamefont {Wu}}, \bibinfo {author} {\bibfnamefont {Y.}~\bibnamefont {Cheng}}, \bibinfo {author} {\bibfnamefont {Z.}~\bibnamefont {Yu}}, \bibinfo {author} {\bibfnamefont {G.}~\bibnamefont {Zhao}},\ and\ \bibinfo {author} {\bibfnamefont {Y.}~\bibnamefont {Chen}},\ }\bibfield  {title} {\bibinfo {title} {Telomere targeting with a novel g-quadruplex-interactive ligand braco-19 induces t-loop disassembly and telomerase displacement in human glioblastoma cells},\ }\href@noop {} {\bibfield  {journal} {\bibinfo  {journal} {Oncotarget}\ }\textbf {\bibinfo {volume} {7}},\ \bibinfo {pages} {14925} (\bibinfo {year} {2016})}\BibitemShut {NoStop}%
\bibitem [{\citenamefont {Neidle}(2010)}]{neidle2010human}%
  \BibitemOpen
  \bibfield  {author} {\bibinfo {author} {\bibfnamefont {S.}~\bibnamefont {Neidle}},\ }\bibfield  {title} {\bibinfo {title} {Human telomeric g-quadruplex: The current status of telomeric g-quadruplexes as therapeutic targets in human cancer},\ }\href@noop {} {\bibfield  {journal} {\bibinfo  {journal} {The FEBS journal}\ }\textbf {\bibinfo {volume} {277}},\ \bibinfo {pages} {1118} (\bibinfo {year} {2010})}\BibitemShut {NoStop}%
\bibitem [{\citenamefont {Perrone}\ \emph {et~al.}(2014)\citenamefont {Perrone}, \citenamefont {Butovskaya}, \citenamefont {Daelemans}, \citenamefont {Palu}, \citenamefont {Pannecouque},\ and\ \citenamefont {Richter}}]{perrone2014anti}%
  \BibitemOpen
  \bibfield  {author} {\bibinfo {author} {\bibfnamefont {R.}~\bibnamefont {Perrone}}, \bibinfo {author} {\bibfnamefont {E.}~\bibnamefont {Butovskaya}}, \bibinfo {author} {\bibfnamefont {D.}~\bibnamefont {Daelemans}}, \bibinfo {author} {\bibfnamefont {G.}~\bibnamefont {Palu}}, \bibinfo {author} {\bibfnamefont {C.}~\bibnamefont {Pannecouque}},\ and\ \bibinfo {author} {\bibfnamefont {S.~N.}\ \bibnamefont {Richter}},\ }\bibfield  {title} {\bibinfo {title} {Anti-hiv-1 activity of the g-quadruplex ligand braco-19},\ }\href@noop {} {\bibfield  {journal} {\bibinfo  {journal} {Journal of antimicrobial chemotherapy}\ }\textbf {\bibinfo {volume} {69}},\ \bibinfo {pages} {3248} (\bibinfo {year} {2014})}\BibitemShut {NoStop}%
\bibitem [{\citenamefont {Yatsunyk}\ \emph {et~al.}(2014)\citenamefont {Yatsunyk}, \citenamefont {Mendoza},\ and\ \citenamefont {Mergny}}]{yatsunyk2014nano}%
  \BibitemOpen
  \bibfield  {author} {\bibinfo {author} {\bibfnamefont {L.~A.}\ \bibnamefont {Yatsunyk}}, \bibinfo {author} {\bibfnamefont {O.}~\bibnamefont {Mendoza}},\ and\ \bibinfo {author} {\bibfnamefont {J.-L.}\ \bibnamefont {Mergny}},\ }\bibfield  {title} {\bibinfo {title} {“nano-oddities”: unusual nucleic acid assemblies for dna-based nanostructures and nanodevices},\ }\href@noop {} {\bibfield  {journal} {\bibinfo  {journal} {Accounts of chemical research}\ }\textbf {\bibinfo {volume} {47}},\ \bibinfo {pages} {1836} (\bibinfo {year} {2014})}\BibitemShut {NoStop}%
\bibitem [{\citenamefont {Mergny}\ and\ \citenamefont {Sen}(2019)}]{mergny2019dna}%
  \BibitemOpen
  \bibfield  {author} {\bibinfo {author} {\bibfnamefont {J.-L.}\ \bibnamefont {Mergny}}\ and\ \bibinfo {author} {\bibfnamefont {D.}~\bibnamefont {Sen}},\ }\bibfield  {title} {\bibinfo {title} {Dna quadruple helices in nanotechnology},\ }\href@noop {} {\bibfield  {journal} {\bibinfo  {journal} {Chemical Reviews}\ }\textbf {\bibinfo {volume} {119}},\ \bibinfo {pages} {6290} (\bibinfo {year} {2019})}\BibitemShut {NoStop}%
\bibitem [{\citenamefont {Kolesnikova}\ and\ \citenamefont {Curtis}(2019)}]{kolesnikova2019structure}%
  \BibitemOpen
  \bibfield  {author} {\bibinfo {author} {\bibfnamefont {S.}~\bibnamefont {Kolesnikova}}\ and\ \bibinfo {author} {\bibfnamefont {E.~A.}\ \bibnamefont {Curtis}},\ }\bibfield  {title} {\bibinfo {title} {Structure and function of multimeric g-quadruplexes},\ }\href@noop {} {\bibfield  {journal} {\bibinfo  {journal} {Molecules}\ }\textbf {\bibinfo {volume} {24}},\ \bibinfo {pages} {3074} (\bibinfo {year} {2019})}\BibitemShut {NoStop}%
\bibitem [{\citenamefont {Frasson}\ \emph {et~al.}(2022)\citenamefont {Frasson}, \citenamefont {Pirota}, \citenamefont {Richter},\ and\ \citenamefont {Doria}}]{frasson2022multimeric}%
  \BibitemOpen
  \bibfield  {author} {\bibinfo {author} {\bibfnamefont {I.}~\bibnamefont {Frasson}}, \bibinfo {author} {\bibfnamefont {V.}~\bibnamefont {Pirota}}, \bibinfo {author} {\bibfnamefont {S.~N.}\ \bibnamefont {Richter}},\ and\ \bibinfo {author} {\bibfnamefont {F.}~\bibnamefont {Doria}},\ }\bibfield  {title} {\bibinfo {title} {Multimeric g-quadruplexes: A review on their biological roles and targeting},\ }\href@noop {} {\bibfield  {journal} {\bibinfo  {journal} {International Journal of Biological Macromolecules}\ }\textbf {\bibinfo {volume} {204}},\ \bibinfo {pages} {89} (\bibinfo {year} {2022})}\BibitemShut {NoStop}%
\bibitem [{\citenamefont {Rigo}\ \emph {et~al.}(2022)\citenamefont {Rigo}, \citenamefont {Groaz},\ and\ \citenamefont {Sissi}}]{rigo2022polymorphic}%
  \BibitemOpen
  \bibfield  {author} {\bibinfo {author} {\bibfnamefont {R.}~\bibnamefont {Rigo}}, \bibinfo {author} {\bibfnamefont {E.}~\bibnamefont {Groaz}},\ and\ \bibinfo {author} {\bibfnamefont {C.}~\bibnamefont {Sissi}},\ }\bibfield  {title} {\bibinfo {title} {Polymorphic and higher-order g-quadruplexes as possible transcription regulators: Novel perspectives for future anticancer therapeutic applications},\ }\href@noop {} {\bibfield  {journal} {\bibinfo  {journal} {Pharmaceuticals}\ }\textbf {\bibinfo {volume} {15}},\ \bibinfo {pages} {373} (\bibinfo {year} {2022})}\BibitemShut {NoStop}%
\bibitem [{\citenamefont {RK}(1988)}]{rk1988highly}%
  \BibitemOpen
  \bibfield  {author} {\bibinfo {author} {\bibfnamefont {M.}~\bibnamefont {RK}},\ }\bibfield  {title} {\bibinfo {title} {A highly conseved repetitive dna sequence,(ttaggg) n, present at the telomeres of human chromosomes},\ }\href@noop {} {\bibfield  {journal} {\bibinfo  {journal} {Proc Natl Acad Sci USA}\ }\textbf {\bibinfo {volume} {85}},\ \bibinfo {pages} {6622} (\bibinfo {year} {1988})}\BibitemShut {NoStop}%
\bibitem [{\citenamefont {Wright}\ \emph {et~al.}(1997)\citenamefont {Wright}, \citenamefont {Tesmer}, \citenamefont {Huffman}, \citenamefont {Levene},\ and\ \citenamefont {Shay}}]{wright1997normal}%
  \BibitemOpen
  \bibfield  {author} {\bibinfo {author} {\bibfnamefont {W.~E.}\ \bibnamefont {Wright}}, \bibinfo {author} {\bibfnamefont {V.~M.}\ \bibnamefont {Tesmer}}, \bibinfo {author} {\bibfnamefont {K.~E.}\ \bibnamefont {Huffman}}, \bibinfo {author} {\bibfnamefont {S.~D.}\ \bibnamefont {Levene}},\ and\ \bibinfo {author} {\bibfnamefont {J.~W.}\ \bibnamefont {Shay}},\ }\bibfield  {title} {\bibinfo {title} {Normal human chromosomes have long g-rich telomeric overhangs at one end},\ }\href@noop {} {\bibfield  {journal} {\bibinfo  {journal} {Genes \& development}\ }\textbf {\bibinfo {volume} {11}},\ \bibinfo {pages} {2801} (\bibinfo {year} {1997})}\BibitemShut {NoStop}%
\bibitem [{\citenamefont {Monsen}\ \emph {et~al.}(2022)\citenamefont {Monsen}, \citenamefont {Trent},\ and\ \citenamefont {Chaires}}]{monsen2022g}%
  \BibitemOpen
  \bibfield  {author} {\bibinfo {author} {\bibfnamefont {R.~C.}\ \bibnamefont {Monsen}}, \bibinfo {author} {\bibfnamefont {J.~O.}\ \bibnamefont {Trent}},\ and\ \bibinfo {author} {\bibfnamefont {J.~B.}\ \bibnamefont {Chaires}},\ }\bibfield  {title} {\bibinfo {title} {G-quadruplex dna: a longer story},\ }\href@noop {} {\bibfield  {journal} {\bibinfo  {journal} {Accounts of chemical research}\ }\textbf {\bibinfo {volume} {55}},\ \bibinfo {pages} {3242} (\bibinfo {year} {2022})}\BibitemShut {NoStop}%
\bibitem [{\citenamefont {Bajpai}\ \emph {et~al.}(2020)\citenamefont {Bajpai}, \citenamefont {Pavlov}, \citenamefont {Lorber}, \citenamefont {Volk},\ and\ \citenamefont {Safran}}]{bajpai2020mesoscale}%
  \BibitemOpen
  \bibfield  {author} {\bibinfo {author} {\bibfnamefont {G.}~\bibnamefont {Bajpai}}, \bibinfo {author} {\bibfnamefont {D.~A.}\ \bibnamefont {Pavlov}}, \bibinfo {author} {\bibfnamefont {D.}~\bibnamefont {Lorber}}, \bibinfo {author} {\bibfnamefont {T.}~\bibnamefont {Volk}},\ and\ \bibinfo {author} {\bibfnamefont {S.}~\bibnamefont {Safran}},\ }\bibfield  {title} {\bibinfo {title} {Mesoscale phase separation of chromatin in the nucleus},\ }\href@noop {} {\bibfield  {journal} {\bibinfo  {journal} {Biophysical Journal}\ }\textbf {\bibinfo {volume} {118}},\ \bibinfo {pages} {549a} (\bibinfo {year} {2020})}\BibitemShut {NoStop}%
\bibitem [{\citenamefont {Aznauryan}\ and\ \citenamefont {Birkedal}(2023)}]{aznauryan2023dynamics}%
  \BibitemOpen
  \bibfield  {author} {\bibinfo {author} {\bibfnamefont {M.}~\bibnamefont {Aznauryan}}\ and\ \bibinfo {author} {\bibfnamefont {V.}~\bibnamefont {Birkedal}},\ }\bibfield  {title} {\bibinfo {title} {Dynamics of g-quadruplex formation under molecular crowding},\ }\href@noop {} {\bibfield  {journal} {\bibinfo  {journal} {The Journal of Physical Chemistry Letters}\ }\textbf {\bibinfo {volume} {14}},\ \bibinfo {pages} {10354} (\bibinfo {year} {2023})}\BibitemShut {NoStop}%
\bibitem [{\citenamefont {Gao}\ \emph {et~al.}(2023)\citenamefont {Gao}, \citenamefont {Mohamed}, \citenamefont {Deng}, \citenamefont {Umer}, \citenamefont {Anwar}, \citenamefont {Chen}, \citenamefont {Wu}, \citenamefont {Wang},\ and\ \citenamefont {He}}]{gao2023effects}%
  \BibitemOpen
  \bibfield  {author} {\bibinfo {author} {\bibfnamefont {C.}~\bibnamefont {Gao}}, \bibinfo {author} {\bibfnamefont {H.~I.}\ \bibnamefont {Mohamed}}, \bibinfo {author} {\bibfnamefont {J.}~\bibnamefont {Deng}}, \bibinfo {author} {\bibfnamefont {M.}~\bibnamefont {Umer}}, \bibinfo {author} {\bibfnamefont {N.}~\bibnamefont {Anwar}}, \bibinfo {author} {\bibfnamefont {J.}~\bibnamefont {Chen}}, \bibinfo {author} {\bibfnamefont {Q.}~\bibnamefont {Wu}}, \bibinfo {author} {\bibfnamefont {Z.}~\bibnamefont {Wang}},\ and\ \bibinfo {author} {\bibfnamefont {Y.}~\bibnamefont {He}},\ }\bibfield  {title} {\bibinfo {title} {Effects of molecular crowding on the structure, stability, and interaction with ligands of g-quadruplexes},\ }\href@noop {} {\bibfield  {journal} {\bibinfo  {journal} {ACS omega}\ }\textbf {\bibinfo {volume} {8}},\ \bibinfo {pages} {14342} (\bibinfo {year} {2023})}\BibitemShut {NoStop}%
\bibitem [{\citenamefont {Yu}\ \emph {et~al.}(2006)\citenamefont {Yu}, \citenamefont {Miyoshi},\ and\ \citenamefont {Sugimoto}}]{yu2006characterization}%
  \BibitemOpen
  \bibfield  {author} {\bibinfo {author} {\bibfnamefont {H.-Q.}\ \bibnamefont {Yu}}, \bibinfo {author} {\bibfnamefont {D.}~\bibnamefont {Miyoshi}},\ and\ \bibinfo {author} {\bibfnamefont {N.}~\bibnamefont {Sugimoto}},\ }\bibfield  {title} {\bibinfo {title} {Characterization of structure and stability of long telomeric dna g-quadruplexes},\ }\href@noop {} {\bibfield  {journal} {\bibinfo  {journal} {Journal of the American Chemical Society}\ }\textbf {\bibinfo {volume} {128}},\ \bibinfo {pages} {15461} (\bibinfo {year} {2006})}\BibitemShut {NoStop}%
\bibitem [{\citenamefont {Ren{\v{c}}iuk}\ \emph {et~al.}(2009)\citenamefont {Ren{\v{c}}iuk}, \citenamefont {Kejnovsk{\'a}}, \citenamefont {{\v{S}}kol{\'a}kov{\'a}}, \citenamefont {Bedn{\'a}{\v{r}}ov{\'a}}, \citenamefont {Motlov{\'a}},\ and\ \citenamefont {Vorl{\'\i}{\v{c}}kov{\'a}}}]{renvciuk2009arrangements}%
  \BibitemOpen
  \bibfield  {author} {\bibinfo {author} {\bibfnamefont {D.}~\bibnamefont {Ren{\v{c}}iuk}}, \bibinfo {author} {\bibfnamefont {I.}~\bibnamefont {Kejnovsk{\'a}}}, \bibinfo {author} {\bibfnamefont {P.}~\bibnamefont {{\v{S}}kol{\'a}kov{\'a}}}, \bibinfo {author} {\bibfnamefont {K.}~\bibnamefont {Bedn{\'a}{\v{r}}ov{\'a}}}, \bibinfo {author} {\bibfnamefont {J.}~\bibnamefont {Motlov{\'a}}},\ and\ \bibinfo {author} {\bibfnamefont {M.}~\bibnamefont {Vorl{\'\i}{\v{c}}kov{\'a}}},\ }\bibfield  {title} {\bibinfo {title} {Arrangements of human telomere dna quadruplex in physiologically relevant k+ solutions},\ }\href@noop {} {\bibfield  {journal} {\bibinfo  {journal} {Nucleic acids research}\ }\textbf {\bibinfo {volume} {37}},\ \bibinfo {pages} {6625} (\bibinfo {year} {2009})}\BibitemShut {NoStop}%
\bibitem [{\citenamefont {Xu}\ \emph {et~al.}(2009)\citenamefont {Xu}, \citenamefont {Ishizuka}, \citenamefont {Kurabayashi},\ and\ \citenamefont {Komiyama}}]{xu2009consecutive}%
  \BibitemOpen
  \bibfield  {author} {\bibinfo {author} {\bibfnamefont {Y.}~\bibnamefont {Xu}}, \bibinfo {author} {\bibfnamefont {T.}~\bibnamefont {Ishizuka}}, \bibinfo {author} {\bibfnamefont {K.}~\bibnamefont {Kurabayashi}},\ and\ \bibinfo {author} {\bibfnamefont {M.}~\bibnamefont {Komiyama}},\ }\bibfield  {title} {\bibinfo {title} {Consecutive formation of g-quadruplexes in human telomeric-overhang dna: a protective capping structure for telomere ends},\ }\href@noop {} {\bibfield  {journal} {\bibinfo  {journal} {Angewandte Chemie}\ }\textbf {\bibinfo {volume} {121}},\ \bibinfo {pages} {7973} (\bibinfo {year} {2009})}\BibitemShut {NoStop}%
\bibitem [{\citenamefont {Abraham~Punnoose}\ \emph {et~al.}(2014)\citenamefont {Abraham~Punnoose}, \citenamefont {Cui}, \citenamefont {Koirala}, \citenamefont {Yangyuoru}, \citenamefont {Ghimire}, \citenamefont {Shrestha},\ and\ \citenamefont {Mao}}]{abraham2014interaction}%
  \BibitemOpen
  \bibfield  {author} {\bibinfo {author} {\bibfnamefont {J.}~\bibnamefont {Abraham~Punnoose}}, \bibinfo {author} {\bibfnamefont {Y.}~\bibnamefont {Cui}}, \bibinfo {author} {\bibfnamefont {D.}~\bibnamefont {Koirala}}, \bibinfo {author} {\bibfnamefont {P.~M.}\ \bibnamefont {Yangyuoru}}, \bibinfo {author} {\bibfnamefont {C.}~\bibnamefont {Ghimire}}, \bibinfo {author} {\bibfnamefont {P.}~\bibnamefont {Shrestha}},\ and\ \bibinfo {author} {\bibfnamefont {H.}~\bibnamefont {Mao}},\ }\bibfield  {title} {\bibinfo {title} {Interaction of g-quadruplexes in the full-length 3 human telomeric overhang},\ }\href@noop {} {\bibfield  {journal} {\bibinfo  {journal} {Journal of the American Chemical Society}\ }\textbf {\bibinfo {volume} {136}},\ \bibinfo {pages} {18062} (\bibinfo {year} {2014})}\BibitemShut {NoStop}%
\bibitem [{\citenamefont {Bugaut}\ and\ \citenamefont {Alberti}(2015)}]{bugaut2015understanding}%
  \BibitemOpen
  \bibfield  {author} {\bibinfo {author} {\bibfnamefont {A.}~\bibnamefont {Bugaut}}\ and\ \bibinfo {author} {\bibfnamefont {P.}~\bibnamefont {Alberti}},\ }\bibfield  {title} {\bibinfo {title} {Understanding the stability of dna g-quadruplex units in long human telomeric strands},\ }\href@noop {} {\bibfield  {journal} {\bibinfo  {journal} {Biochimie}\ }\textbf {\bibinfo {volume} {113}},\ \bibinfo {pages} {125} (\bibinfo {year} {2015})}\BibitemShut {NoStop}%
\bibitem [{\citenamefont {Petraccone}\ \emph {et~al.}(2010)\citenamefont {Petraccone}, \citenamefont {Garbett}, \citenamefont {Chaires},\ and\ \citenamefont {Trent}}]{petraccone2010integrated}%
  \BibitemOpen
  \bibfield  {author} {\bibinfo {author} {\bibfnamefont {L.}~\bibnamefont {Petraccone}}, \bibinfo {author} {\bibfnamefont {N.~C.}\ \bibnamefont {Garbett}}, \bibinfo {author} {\bibfnamefont {J.~B.}\ \bibnamefont {Chaires}},\ and\ \bibinfo {author} {\bibfnamefont {J.~O.}\ \bibnamefont {Trent}},\ }\bibfield  {title} {\bibinfo {title} {An integrated molecular dynamics (md) and experimental study of higher order human telomeric quadruplexes},\ }\href@noop {} {\bibfield  {journal} {\bibinfo  {journal} {Biopolymers: Original Research on Biomolecules}\ }\textbf {\bibinfo {volume} {93}},\ \bibinfo {pages} {533} (\bibinfo {year} {2010})}\BibitemShut {NoStop}%
\bibitem [{\citenamefont {Petraccone}\ \emph {et~al.}(2011)\citenamefont {Petraccone}, \citenamefont {Spink}, \citenamefont {Trent}, \citenamefont {Garbett}, \citenamefont {Mekmaysy}, \citenamefont {Giancola},\ and\ \citenamefont {Chaires}}]{petraccone2011structure}%
  \BibitemOpen
  \bibfield  {author} {\bibinfo {author} {\bibfnamefont {L.}~\bibnamefont {Petraccone}}, \bibinfo {author} {\bibfnamefont {C.}~\bibnamefont {Spink}}, \bibinfo {author} {\bibfnamefont {J.~O.}\ \bibnamefont {Trent}}, \bibinfo {author} {\bibfnamefont {N.~C.}\ \bibnamefont {Garbett}}, \bibinfo {author} {\bibfnamefont {C.~S.}\ \bibnamefont {Mekmaysy}}, \bibinfo {author} {\bibfnamefont {C.}~\bibnamefont {Giancola}},\ and\ \bibinfo {author} {\bibfnamefont {J.~B.}\ \bibnamefont {Chaires}},\ }\bibfield  {title} {\bibinfo {title} {Structure and stability of higher-order human telomeric quadruplexes},\ }\href@noop {} {\bibfield  {journal} {\bibinfo  {journal} {Journal of the American Chemical Society}\ }\textbf {\bibinfo {volume} {133}},\ \bibinfo {pages} {20951} (\bibinfo {year} {2011})}\BibitemShut {NoStop}%
\bibitem [{\citenamefont {Chaires}\ \emph {et~al.}(2015)\citenamefont {Chaires}, \citenamefont {Dean}, \citenamefont {Le},\ and\ \citenamefont {Trent}}]{chaires2015hydrodynamic}%
  \BibitemOpen
  \bibfield  {author} {\bibinfo {author} {\bibfnamefont {J.~B.}\ \bibnamefont {Chaires}}, \bibinfo {author} {\bibfnamefont {W.~L.}\ \bibnamefont {Dean}}, \bibinfo {author} {\bibfnamefont {H.~T.}\ \bibnamefont {Le}},\ and\ \bibinfo {author} {\bibfnamefont {J.~O.}\ \bibnamefont {Trent}},\ }\bibfield  {title} {\bibinfo {title} {Hydrodynamic models of g-quadruplex structures},\ }in\ \href@noop {} {\emph {\bibinfo {booktitle} {Methods in enzymology}}},\ Vol.\ \bibinfo {volume} {562}\ (\bibinfo  {publisher} {Elsevier},\ \bibinfo {year} {2015})\ pp.\ \bibinfo {pages} {287--304}\BibitemShut {NoStop}%
\bibitem [{\citenamefont {Wang}\ \emph {et~al.}(2011)\citenamefont {Wang}, \citenamefont {Nora}, \citenamefont {Ghodke},\ and\ \citenamefont {Opresko}}]{wang2011single}%
  \BibitemOpen
  \bibfield  {author} {\bibinfo {author} {\bibfnamefont {H.}~\bibnamefont {Wang}}, \bibinfo {author} {\bibfnamefont {G.~J.}\ \bibnamefont {Nora}}, \bibinfo {author} {\bibfnamefont {H.}~\bibnamefont {Ghodke}},\ and\ \bibinfo {author} {\bibfnamefont {P.~L.}\ \bibnamefont {Opresko}},\ }\bibfield  {title} {\bibinfo {title} {Single molecule studies of physiologically relevant telomeric tails reveal pot1 mechanism for promoting g-quadruplex unfolding},\ }\href@noop {} {\bibfield  {journal} {\bibinfo  {journal} {Journal of Biological Chemistry}\ }\textbf {\bibinfo {volume} {286}},\ \bibinfo {pages} {7479} (\bibinfo {year} {2011})}\BibitemShut {NoStop}%
\bibitem [{\citenamefont {Kar}\ \emph {et~al.}(2018)\citenamefont {Kar}, \citenamefont {Jones}, \citenamefont {Arat}, \citenamefont {Fishel},\ and\ \citenamefont {Griffith}}]{kar2018long}%
  \BibitemOpen
  \bibfield  {author} {\bibinfo {author} {\bibfnamefont {A.}~\bibnamefont {Kar}}, \bibinfo {author} {\bibfnamefont {N.}~\bibnamefont {Jones}}, \bibinfo {author} {\bibfnamefont {N.~{\"O}.}\ \bibnamefont {Arat}}, \bibinfo {author} {\bibfnamefont {R.}~\bibnamefont {Fishel}},\ and\ \bibinfo {author} {\bibfnamefont {J.~D.}\ \bibnamefont {Griffith}},\ }\bibfield  {title} {\bibinfo {title} {Long repeating (ttaggg) n single-stranded dna self-condenses into compact beaded filaments stabilized by g-quadruplex formation},\ }\href@noop {} {\bibfield  {journal} {\bibinfo  {journal} {Journal of Biological Chemistry}\ }\textbf {\bibinfo {volume} {293}},\ \bibinfo {pages} {9473} (\bibinfo {year} {2018})}\BibitemShut {NoStop}%
\bibitem [{\citenamefont {Abraham~Punnoose}\ \emph {et~al.}(2018)\citenamefont {Abraham~Punnoose}, \citenamefont {Ma}, \citenamefont {Hoque}, \citenamefont {Cui}, \citenamefont {Sasaki}, \citenamefont {Guo}, \citenamefont {Nagasawa},\ and\ \citenamefont {Mao}}]{abraham2018random}%
  \BibitemOpen
  \bibfield  {author} {\bibinfo {author} {\bibfnamefont {J.}~\bibnamefont {Abraham~Punnoose}}, \bibinfo {author} {\bibfnamefont {Y.}~\bibnamefont {Ma}}, \bibinfo {author} {\bibfnamefont {M.~E.}\ \bibnamefont {Hoque}}, \bibinfo {author} {\bibfnamefont {Y.}~\bibnamefont {Cui}}, \bibinfo {author} {\bibfnamefont {S.}~\bibnamefont {Sasaki}}, \bibinfo {author} {\bibfnamefont {A.~H.}\ \bibnamefont {Guo}}, \bibinfo {author} {\bibfnamefont {K.}~\bibnamefont {Nagasawa}},\ and\ \bibinfo {author} {\bibfnamefont {H.}~\bibnamefont {Mao}},\ }\bibfield  {title} {\bibinfo {title} {Random formation of g-quadruplexes in the full-length human telomere overhangs leads to a kinetic folding pattern with targetable vacant g-tracts},\ }\href@noop {} {\bibfield  {journal} {\bibinfo  {journal} {Biochemistry}\ }\textbf {\bibinfo {volume} {57}},\ \bibinfo {pages} {6946} (\bibinfo {year} {2018})}\BibitemShut {NoStop}%
\bibitem [{\citenamefont {Henzler-Wildman}\ and\ \citenamefont {Kern}(2007)}]{henzler2007dynamic}%
  \BibitemOpen
  \bibfield  {author} {\bibinfo {author} {\bibfnamefont {K.}~\bibnamefont {Henzler-Wildman}}\ and\ \bibinfo {author} {\bibfnamefont {D.}~\bibnamefont {Kern}},\ }\bibfield  {title} {\bibinfo {title} {Dynamic personalities of proteins},\ }\href@noop {} {\bibfield  {journal} {\bibinfo  {journal} {Nature}\ }\textbf {\bibinfo {volume} {450}},\ \bibinfo {pages} {964} (\bibinfo {year} {2007})}\BibitemShut {NoStop}%
\bibitem [{\citenamefont {Karplus}\ and\ \citenamefont {Kuriyan}(2005)}]{karplus2005molecular}%
  \BibitemOpen
  \bibfield  {author} {\bibinfo {author} {\bibfnamefont {M.}~\bibnamefont {Karplus}}\ and\ \bibinfo {author} {\bibfnamefont {J.}~\bibnamefont {Kuriyan}},\ }\bibfield  {title} {\bibinfo {title} {Molecular dynamics and protein function},\ }\href@noop {} {\bibfield  {journal} {\bibinfo  {journal} {Proceedings of the National Academy of Sciences}\ }\textbf {\bibinfo {volume} {102}},\ \bibinfo {pages} {6679} (\bibinfo {year} {2005})}\BibitemShut {NoStop}%
\bibitem [{\citenamefont {Zimmerberg}\ and\ \citenamefont {Kozlov}(2006)}]{zimmerberg2006proteins}%
  \BibitemOpen
  \bibfield  {author} {\bibinfo {author} {\bibfnamefont {J.}~\bibnamefont {Zimmerberg}}\ and\ \bibinfo {author} {\bibfnamefont {M.~M.}\ \bibnamefont {Kozlov}},\ }\bibfield  {title} {\bibinfo {title} {How proteins produce cellular membrane curvature},\ }\href@noop {} {\bibfield  {journal} {\bibinfo  {journal} {Nature reviews Molecular cell biology}\ }\textbf {\bibinfo {volume} {7}},\ \bibinfo {pages} {9} (\bibinfo {year} {2006})}\BibitemShut {NoStop}%
\bibitem [{\citenamefont {Pyle}(2002)}]{pyle2002metal}%
  \BibitemOpen
  \bibfield  {author} {\bibinfo {author} {\bibfnamefont {A.}~\bibnamefont {Pyle}},\ }\bibfield  {title} {\bibinfo {title} {Metal ions in the structure and function of rna},\ }\href@noop {} {\bibfield  {journal} {\bibinfo  {journal} {JBIC Journal of Biological Inorganic Chemistry}\ }\textbf {\bibinfo {volume} {7}},\ \bibinfo {pages} {679} (\bibinfo {year} {2002})}\BibitemShut {NoStop}%
\bibitem [{\citenamefont {Hsu}\ \emph {et~al.}(2009)\citenamefont {Hsu}, \citenamefont {Binder},\ and\ \citenamefont {Paul}}]{hsu2009define}%
  \BibitemOpen
  \bibfield  {author} {\bibinfo {author} {\bibfnamefont {H.-P.}\ \bibnamefont {Hsu}}, \bibinfo {author} {\bibfnamefont {K.}~\bibnamefont {Binder}},\ and\ \bibinfo {author} {\bibfnamefont {W.}~\bibnamefont {Paul}},\ }\bibfield  {title} {\bibinfo {title} {How to define variation of physical properties normal to an undulating one-dimensional object},\ }\href@noop {} {\bibfield  {journal} {\bibinfo  {journal} {Physical review letters}\ }\textbf {\bibinfo {volume} {103}},\ \bibinfo {pages} {198301} (\bibinfo {year} {2009})}\BibitemShut {NoStop}%
\bibitem [{\citenamefont {Zhao}\ and\ \citenamefont {Zhai}(2020)}]{zhao2020recent}%
  \BibitemOpen
  \bibfield  {author} {\bibinfo {author} {\bibfnamefont {J.}~\bibnamefont {Zhao}}\ and\ \bibinfo {author} {\bibfnamefont {Q.}~\bibnamefont {Zhai}},\ }\bibfield  {title} {\bibinfo {title} {Recent advances in the development of ligands specifically targeting telomeric multimeric g-quadruplexes},\ }\href@noop {} {\bibfield  {journal} {\bibinfo  {journal} {Bioorganic Chemistry}\ }\textbf {\bibinfo {volume} {103}},\ \bibinfo {pages} {104229} (\bibinfo {year} {2020})}\BibitemShut {NoStop}%
\bibitem [{\citenamefont {Figueiredo}\ \emph {et~al.}(2024)\citenamefont {Figueiredo}, \citenamefont {Mergny},\ and\ \citenamefont {Cruz}}]{figueiredo2024g}%
  \BibitemOpen
  \bibfield  {author} {\bibinfo {author} {\bibfnamefont {J.}~\bibnamefont {Figueiredo}}, \bibinfo {author} {\bibfnamefont {J.-L.}\ \bibnamefont {Mergny}},\ and\ \bibinfo {author} {\bibfnamefont {C.}~\bibnamefont {Cruz}},\ }\bibfield  {title} {\bibinfo {title} {G-quadruplex ligands in cancer therapy: Progress, challenges, and clinical perspectives},\ }\href@noop {} {\bibfield  {journal} {\bibinfo  {journal} {Life Sciences}\ ,\ \bibinfo {pages} {122481}} (\bibinfo {year} {2024})}\BibitemShut {NoStop}%
\bibitem [{\citenamefont {Manning}(1969)}]{manning1969limiting}%
  \BibitemOpen
  \bibfield  {author} {\bibinfo {author} {\bibfnamefont {G.~S.}\ \bibnamefont {Manning}},\ }\bibfield  {title} {\bibinfo {title} {Limiting laws and counterion condensation in polyelectrolyte solutions i. colligative properties},\ }\href@noop {} {\bibfield  {journal} {\bibinfo  {journal} {The journal of chemical Physics}\ }\textbf {\bibinfo {volume} {51}},\ \bibinfo {pages} {924} (\bibinfo {year} {1969})}\BibitemShut {NoStop}%
\bibitem [{\citenamefont {Monsen}\ \emph {et~al.}(2021)\citenamefont {Monsen}, \citenamefont {Chakravarthy}, \citenamefont {Dean}, \citenamefont {Chaires},\ and\ \citenamefont {Trent}}]{monsen2021solution}%
  \BibitemOpen
  \bibfield  {author} {\bibinfo {author} {\bibfnamefont {R.~C.}\ \bibnamefont {Monsen}}, \bibinfo {author} {\bibfnamefont {S.}~\bibnamefont {Chakravarthy}}, \bibinfo {author} {\bibfnamefont {W.~L.}\ \bibnamefont {Dean}}, \bibinfo {author} {\bibfnamefont {J.~B.}\ \bibnamefont {Chaires}},\ and\ \bibinfo {author} {\bibfnamefont {J.~O.}\ \bibnamefont {Trent}},\ }\bibfield  {title} {\bibinfo {title} {The solution structures of higher-order human telomere g-quadruplex multimers},\ }\href@noop {} {\bibfield  {journal} {\bibinfo  {journal} {Nucleic Acids Research}\ }\textbf {\bibinfo {volume} {49}},\ \bibinfo {pages} {1749} (\bibinfo {year} {2021})}\BibitemShut {NoStop}%
\bibitem [{\citenamefont {Limongelli}\ \emph {et~al.}(2013)\citenamefont {Limongelli}, \citenamefont {De~Tito}, \citenamefont {Cerofolini}, \citenamefont {Fragai}, \citenamefont {Pagano}, \citenamefont {Trotta}, \citenamefont {Cosconati}, \citenamefont {Marinelli}, \citenamefont {Novellino}, \citenamefont {Bertini} \emph {et~al.}}]{limongelli2013g}%
  \BibitemOpen
  \bibfield  {author} {\bibinfo {author} {\bibfnamefont {V.}~\bibnamefont {Limongelli}}, \bibinfo {author} {\bibfnamefont {S.}~\bibnamefont {De~Tito}}, \bibinfo {author} {\bibfnamefont {L.}~\bibnamefont {Cerofolini}}, \bibinfo {author} {\bibfnamefont {M.}~\bibnamefont {Fragai}}, \bibinfo {author} {\bibfnamefont {B.}~\bibnamefont {Pagano}}, \bibinfo {author} {\bibfnamefont {R.}~\bibnamefont {Trotta}}, \bibinfo {author} {\bibfnamefont {S.}~\bibnamefont {Cosconati}}, \bibinfo {author} {\bibfnamefont {L.}~\bibnamefont {Marinelli}}, \bibinfo {author} {\bibfnamefont {E.}~\bibnamefont {Novellino}}, \bibinfo {author} {\bibfnamefont {I.}~\bibnamefont {Bertini}}, \emph {et~al.},\ }\bibfield  {title} {\bibinfo {title} {The g-triplex dna},\ }\href@noop {} {\bibfield  {journal} {\bibinfo  {journal} {Angewandte Chemie International Edition}\ }\textbf {\bibinfo {volume} {52}},\ \bibinfo {pages} {2269} (\bibinfo {year} {2013})}\BibitemShut {NoStop}%
\bibitem [{\citenamefont {Galindo-Murillo}\ \emph {et~al.}(2016)\citenamefont {Galindo-Murillo}, \citenamefont {Robertson}, \citenamefont {Zgarbova}, \citenamefont {Sponer}, \citenamefont {Otyepka}, \citenamefont {Jurecka},\ and\ \citenamefont {Cheatham~III}}]{galindo2016assessing}%
  \BibitemOpen
  \bibfield  {author} {\bibinfo {author} {\bibfnamefont {R.}~\bibnamefont {Galindo-Murillo}}, \bibinfo {author} {\bibfnamefont {J.~C.}\ \bibnamefont {Robertson}}, \bibinfo {author} {\bibfnamefont {M.}~\bibnamefont {Zgarbova}}, \bibinfo {author} {\bibfnamefont {J.}~\bibnamefont {Sponer}}, \bibinfo {author} {\bibfnamefont {M.}~\bibnamefont {Otyepka}}, \bibinfo {author} {\bibfnamefont {P.}~\bibnamefont {Jurecka}},\ and\ \bibinfo {author} {\bibfnamefont {T.~E.}\ \bibnamefont {Cheatham~III}},\ }\bibfield  {title} {\bibinfo {title} {Assessing the current state of amber force field modifications for dna},\ }\href@noop {} {\bibfield  {journal} {\bibinfo  {journal} {Journal of chemical theory and computation}\ }\textbf {\bibinfo {volume} {12}},\ \bibinfo {pages} {4114} (\bibinfo {year} {2016})}\BibitemShut {NoStop}%
\bibitem [{\citenamefont {Islam}\ \emph {et~al.}(2013)\citenamefont {Islam}, \citenamefont {Sgobba}, \citenamefont {Laughton}, \citenamefont {Orozco}, \citenamefont {Sponer}, \citenamefont {Neidle},\ and\ \citenamefont {Haider}}]{islam2013conformational}%
  \BibitemOpen
  \bibfield  {author} {\bibinfo {author} {\bibfnamefont {B.}~\bibnamefont {Islam}}, \bibinfo {author} {\bibfnamefont {M.}~\bibnamefont {Sgobba}}, \bibinfo {author} {\bibfnamefont {C.}~\bibnamefont {Laughton}}, \bibinfo {author} {\bibfnamefont {M.}~\bibnamefont {Orozco}}, \bibinfo {author} {\bibfnamefont {J.}~\bibnamefont {Sponer}}, \bibinfo {author} {\bibfnamefont {S.}~\bibnamefont {Neidle}},\ and\ \bibinfo {author} {\bibfnamefont {S.}~\bibnamefont {Haider}},\ }\bibfield  {title} {\bibinfo {title} {Conformational dynamics of the human propeller telomeric dna quadruplex on a microsecond time scale},\ }\href@noop {} {\bibfield  {journal} {\bibinfo  {journal} {Nucleic acids research}\ }\textbf {\bibinfo {volume} {41}},\ \bibinfo {pages} {2723} (\bibinfo {year} {2013})}\BibitemShut {NoStop}%
\bibitem [{\citenamefont {Wei}\ \emph {et~al.}(2015)\citenamefont {Wei}, \citenamefont {Husby},\ and\ \citenamefont {Neidle}}]{wei2015flexibility}%
  \BibitemOpen
  \bibfield  {author} {\bibinfo {author} {\bibfnamefont {D.}~\bibnamefont {Wei}}, \bibinfo {author} {\bibfnamefont {J.}~\bibnamefont {Husby}},\ and\ \bibinfo {author} {\bibfnamefont {S.}~\bibnamefont {Neidle}},\ }\bibfield  {title} {\bibinfo {title} {Flexibility and structural conservation in a c-kit g-quadruplex},\ }\href@noop {} {\bibfield  {journal} {\bibinfo  {journal} {Nucleic acids research}\ }\textbf {\bibinfo {volume} {43}},\ \bibinfo {pages} {629} (\bibinfo {year} {2015})}\BibitemShut {NoStop}%
\bibitem [{\citenamefont {Ghosh}\ \emph {et~al.}(2015)\citenamefont {Ghosh}, \citenamefont {Jana}, \citenamefont {Kar}, \citenamefont {Chatterjee},\ and\ \citenamefont {Dasgupta}}]{ghosh2015plant}%
  \BibitemOpen
  \bibfield  {author} {\bibinfo {author} {\bibfnamefont {S.}~\bibnamefont {Ghosh}}, \bibinfo {author} {\bibfnamefont {J.}~\bibnamefont {Jana}}, \bibinfo {author} {\bibfnamefont {R.~K.}\ \bibnamefont {Kar}}, \bibinfo {author} {\bibfnamefont {S.}~\bibnamefont {Chatterjee}},\ and\ \bibinfo {author} {\bibfnamefont {D.}~\bibnamefont {Dasgupta}},\ }\bibfield  {title} {\bibinfo {title} {Plant alkaloid chelerythrine induced aggregation of human telomere sequence a unique mode of association between a small molecule and a quadruplex},\ }\href@noop {} {\bibfield  {journal} {\bibinfo  {journal} {Biochemistry}\ }\textbf {\bibinfo {volume} {54}},\ \bibinfo {pages} {974} (\bibinfo {year} {2015})}\BibitemShut {NoStop}%
\bibitem [{\citenamefont {Bian}\ \emph {et~al.}(2014)\citenamefont {Bian}, \citenamefont {Tan}, \citenamefont {Wang}, \citenamefont {Sheng}, \citenamefont {Zhang},\ and\ \citenamefont {Wang}}]{bian2014atomistic}%
  \BibitemOpen
  \bibfield  {author} {\bibinfo {author} {\bibfnamefont {Y.}~\bibnamefont {Bian}}, \bibinfo {author} {\bibfnamefont {C.}~\bibnamefont {Tan}}, \bibinfo {author} {\bibfnamefont {J.}~\bibnamefont {Wang}}, \bibinfo {author} {\bibfnamefont {Y.}~\bibnamefont {Sheng}}, \bibinfo {author} {\bibfnamefont {J.}~\bibnamefont {Zhang}},\ and\ \bibinfo {author} {\bibfnamefont {W.}~\bibnamefont {Wang}},\ }\bibfield  {title} {\bibinfo {title} {Atomistic picture for the folding pathway of a hybrid-1 type human telomeric dna g-quadruplex},\ }\href@noop {} {\bibfield  {journal} {\bibinfo  {journal} {PLoS computational biology}\ }\textbf {\bibinfo {volume} {10}},\ \bibinfo {pages} {e1003562} (\bibinfo {year} {2014})}\BibitemShut {NoStop}%
\bibitem [{\citenamefont {Yang}\ \emph {et~al.}(2017)\citenamefont {Yang}, \citenamefont {Kulkarni}, \citenamefont {Lim},\ and\ \citenamefont {Pak}}]{yang2017silico}%
  \BibitemOpen
  \bibfield  {author} {\bibinfo {author} {\bibfnamefont {C.}~\bibnamefont {Yang}}, \bibinfo {author} {\bibfnamefont {M.}~\bibnamefont {Kulkarni}}, \bibinfo {author} {\bibfnamefont {M.}~\bibnamefont {Lim}},\ and\ \bibinfo {author} {\bibfnamefont {Y.}~\bibnamefont {Pak}},\ }\bibfield  {title} {\bibinfo {title} {In silico direct folding of thrombin-binding aptamer g-quadruplex at all-atom level},\ }\href@noop {} {\bibfield  {journal} {\bibinfo  {journal} {Nucleic acids research}\ }\textbf {\bibinfo {volume} {45}},\ \bibinfo {pages} {12648} (\bibinfo {year} {2017})}\BibitemShut {NoStop}%
\bibitem [{\citenamefont {Stadlbauer}\ \emph {et~al.}(2019)\citenamefont {Stadlbauer}, \citenamefont {K{\"u}hrov{\'a}}, \citenamefont {Vicherek}, \citenamefont {Ban{\'a}{\v{s}}}, \citenamefont {Otyepka}, \citenamefont {Trant{\'\i}rek},\ and\ \citenamefont {{\v{S}}poner}}]{stadlbauer2019parallel}%
  \BibitemOpen
  \bibfield  {author} {\bibinfo {author} {\bibfnamefont {P.}~\bibnamefont {Stadlbauer}}, \bibinfo {author} {\bibfnamefont {P.}~\bibnamefont {K{\"u}hrov{\'a}}}, \bibinfo {author} {\bibfnamefont {L.}~\bibnamefont {Vicherek}}, \bibinfo {author} {\bibfnamefont {P.}~\bibnamefont {Ban{\'a}{\v{s}}}}, \bibinfo {author} {\bibfnamefont {M.}~\bibnamefont {Otyepka}}, \bibinfo {author} {\bibfnamefont {L.}~\bibnamefont {Trant{\'\i}rek}},\ and\ \bibinfo {author} {\bibfnamefont {J.}~\bibnamefont {{\v{S}}poner}},\ }\bibfield  {title} {\bibinfo {title} {Parallel g-triplexes and g-hairpins as potential transitory ensembles in the folding of parallel-stranded dna g-quadruplexes},\ }\href@noop {} {\bibfield  {journal} {\bibinfo  {journal} {Nucleic acids research}\ }\textbf {\bibinfo {volume} {47}},\ \bibinfo {pages} {7276} (\bibinfo {year} {2019})}\BibitemShut {NoStop}%
\bibitem [{\citenamefont {Rocca}\ \emph {et~al.}(2020)\citenamefont {Rocca}, \citenamefont {Palazzesi}, \citenamefont {Amato}, \citenamefont {Costa}, \citenamefont {Ortuso}, \citenamefont {Pagano}, \citenamefont {Randazzo}, \citenamefont {Novellino}, \citenamefont {Alcaro}, \citenamefont {Moraca} \emph {et~al.}}]{rocca2020folding}%
  \BibitemOpen
  \bibfield  {author} {\bibinfo {author} {\bibfnamefont {R.}~\bibnamefont {Rocca}}, \bibinfo {author} {\bibfnamefont {F.}~\bibnamefont {Palazzesi}}, \bibinfo {author} {\bibfnamefont {J.}~\bibnamefont {Amato}}, \bibinfo {author} {\bibfnamefont {G.}~\bibnamefont {Costa}}, \bibinfo {author} {\bibfnamefont {F.}~\bibnamefont {Ortuso}}, \bibinfo {author} {\bibfnamefont {B.}~\bibnamefont {Pagano}}, \bibinfo {author} {\bibfnamefont {A.}~\bibnamefont {Randazzo}}, \bibinfo {author} {\bibfnamefont {E.}~\bibnamefont {Novellino}}, \bibinfo {author} {\bibfnamefont {S.}~\bibnamefont {Alcaro}}, \bibinfo {author} {\bibfnamefont {F.}~\bibnamefont {Moraca}}, \emph {et~al.},\ }\bibfield  {title} {\bibinfo {title} {Folding intermediate states of the parallel human telomeric g-quadruplex dna explored using well-tempered metadynamics},\ }\href@noop {} {\bibfield  {journal} {\bibinfo  {journal} {Scientific reports}\ }\textbf {\bibinfo {volume} {10}},\ \bibinfo {pages} {3176} (\bibinfo {year} {2020})}\BibitemShut {NoStop}%
\bibitem [{\citenamefont {Pokorn{\'a}}\ \emph {et~al.}(2024)\citenamefont {Pokorn{\'a}}, \citenamefont {Ml{\`y}nsk{\`y}}, \citenamefont {Bussi}, \citenamefont {{\v{S}}poner},\ and\ \citenamefont {Stadlbauer}}]{pokorna2024molecular}%
  \BibitemOpen
  \bibfield  {author} {\bibinfo {author} {\bibfnamefont {P.}~\bibnamefont {Pokorn{\'a}}}, \bibinfo {author} {\bibfnamefont {V.}~\bibnamefont {Ml{\`y}nsk{\`y}}}, \bibinfo {author} {\bibfnamefont {G.}~\bibnamefont {Bussi}}, \bibinfo {author} {\bibfnamefont {J.}~\bibnamefont {{\v{S}}poner}},\ and\ \bibinfo {author} {\bibfnamefont {P.}~\bibnamefont {Stadlbauer}},\ }\bibfield  {title} {\bibinfo {title} {Molecular dynamics simulations reveal the parallel stranded d (ggga) 3ggg dna quadruplex folds via multiple paths from a coil-like ensemble},\ }\href@noop {} {\bibfield  {journal} {\bibinfo  {journal} {International Journal of Biological Macromolecules}\ }\textbf {\bibinfo {volume} {261}},\ \bibinfo {pages} {129712} (\bibinfo {year} {2024})}\BibitemShut {NoStop}%
\bibitem [{\citenamefont {Kim}\ \emph {et~al.}(2012)\citenamefont {Kim}, \citenamefont {Yang},\ and\ \citenamefont {Pak}}]{kim2012free}%
  \BibitemOpen
  \bibfield  {author} {\bibinfo {author} {\bibfnamefont {E.}~\bibnamefont {Kim}}, \bibinfo {author} {\bibfnamefont {C.}~\bibnamefont {Yang}},\ and\ \bibinfo {author} {\bibfnamefont {Y.}~\bibnamefont {Pak}},\ }\bibfield  {title} {\bibinfo {title} {Free-energy landscape of a thrombin-binding dna aptamer in aqueous environment},\ }\href@noop {} {\bibfield  {journal} {\bibinfo  {journal} {Journal of chemical theory and computation}\ }\textbf {\bibinfo {volume} {8}},\ \bibinfo {pages} {4845} (\bibinfo {year} {2012})}\BibitemShut {NoStop}%
\bibitem [{\citenamefont {Bergues-Pupo}\ \emph {et~al.}(2015)\citenamefont {Bergues-Pupo}, \citenamefont {Arias-Gonzalez}, \citenamefont {Mor{\'o}n}, \citenamefont {Fiasconaro},\ and\ \citenamefont {Falo}}]{bergues2015role}%
  \BibitemOpen
  \bibfield  {author} {\bibinfo {author} {\bibfnamefont {A.~E.}\ \bibnamefont {Bergues-Pupo}}, \bibinfo {author} {\bibfnamefont {J.~R.}\ \bibnamefont {Arias-Gonzalez}}, \bibinfo {author} {\bibfnamefont {M.~C.}\ \bibnamefont {Mor{\'o}n}}, \bibinfo {author} {\bibfnamefont {A.}~\bibnamefont {Fiasconaro}},\ and\ \bibinfo {author} {\bibfnamefont {F.}~\bibnamefont {Falo}},\ }\bibfield  {title} {\bibinfo {title} {Role of the central cations in the mechanical unfolding of dna and rna g-quadruplexes},\ }\href@noop {} {\bibfield  {journal} {\bibinfo  {journal} {Nucleic acids research}\ }\textbf {\bibinfo {volume} {43}},\ \bibinfo {pages} {7638} (\bibinfo {year} {2015})}\BibitemShut {NoStop}%
\bibitem [{\citenamefont {Zeng}\ \emph {et~al.}(2016)\citenamefont {Zeng}, \citenamefont {Zhang}, \citenamefont {Xiao}, \citenamefont {Jiang}, \citenamefont {Guo}, \citenamefont {Yu}, \citenamefont {Pu},\ and\ \citenamefont {Li}}]{zeng2016unfolding}%
  \BibitemOpen
  \bibfield  {author} {\bibinfo {author} {\bibfnamefont {X.}~\bibnamefont {Zeng}}, \bibinfo {author} {\bibfnamefont {L.}~\bibnamefont {Zhang}}, \bibinfo {author} {\bibfnamefont {X.}~\bibnamefont {Xiao}}, \bibinfo {author} {\bibfnamefont {Y.}~\bibnamefont {Jiang}}, \bibinfo {author} {\bibfnamefont {Y.}~\bibnamefont {Guo}}, \bibinfo {author} {\bibfnamefont {X.}~\bibnamefont {Yu}}, \bibinfo {author} {\bibfnamefont {X.}~\bibnamefont {Pu}},\ and\ \bibinfo {author} {\bibfnamefont {M.}~\bibnamefont {Li}},\ }\bibfield  {title} {\bibinfo {title} {Unfolding mechanism of thrombin-binding aptamer revealed by molecular dynamics simulation and markov state model},\ }\href@noop {} {\bibfield  {journal} {\bibinfo  {journal} {Scientific reports}\ }\textbf {\bibinfo {volume} {6}},\ \bibinfo {pages} {24065} (\bibinfo {year} {2016})}\BibitemShut {NoStop}%
\bibitem [{\citenamefont {Luo}\ and\ \citenamefont {Mu}(2016)}]{luo2016computational}%
  \BibitemOpen
  \bibfield  {author} {\bibinfo {author} {\bibfnamefont {D.}~\bibnamefont {Luo}}\ and\ \bibinfo {author} {\bibfnamefont {Y.}~\bibnamefont {Mu}},\ }\bibfield  {title} {\bibinfo {title} {Computational insights into the stability and folding pathways of human telomeric dna g-quadruplexes},\ }\href@noop {} {\bibfield  {journal} {\bibinfo  {journal} {The Journal of Physical Chemistry B}\ }\textbf {\bibinfo {volume} {120}},\ \bibinfo {pages} {4912} (\bibinfo {year} {2016})}\BibitemShut {NoStop}%
\bibitem [{\citenamefont {Kogut}\ \emph {et~al.}(2016)\citenamefont {Kogut}, \citenamefont {Kleist},\ and\ \citenamefont {Czub}}]{kogut2016molecular}%
  \BibitemOpen
  \bibfield  {author} {\bibinfo {author} {\bibfnamefont {M.}~\bibnamefont {Kogut}}, \bibinfo {author} {\bibfnamefont {C.}~\bibnamefont {Kleist}},\ and\ \bibinfo {author} {\bibfnamefont {J.}~\bibnamefont {Czub}},\ }\bibfield  {title} {\bibinfo {title} {Molecular dynamics simulations reveal the balance of forces governing the formation of a guanine tetrad—a common structural unit of g-quadruplex dna},\ }\href@noop {} {\bibfield  {journal} {\bibinfo  {journal} {Nucleic acids research}\ }\textbf {\bibinfo {volume} {44}},\ \bibinfo {pages} {3020} (\bibinfo {year} {2016})}\BibitemShut {NoStop}%
\bibitem [{\citenamefont {Gajarský}\ \emph {et~al.}(2017)\citenamefont {Gajarský}, \citenamefont {Zivkovic}, \citenamefont {Stadlbauer}, \citenamefont {Pagano}, \citenamefont {Fiala}, \citenamefont {Amato}, \citenamefont {Tomáška}, \citenamefont {Sponer}, \citenamefont {Plavec},\ and\ \citenamefont {Trantirek}}]{gajarský2017structure}%
  \BibitemOpen
  \bibfield  {author} {\bibinfo {author} {\bibfnamefont {M.}~\bibnamefont {Gajarský}}, \bibinfo {author} {\bibfnamefont {M.~L.}\ \bibnamefont {Zivkovic}}, \bibinfo {author} {\bibfnamefont {P.}~\bibnamefont {Stadlbauer}}, \bibinfo {author} {\bibfnamefont {B.}~\bibnamefont {Pagano}}, \bibinfo {author} {\bibfnamefont {R.}~\bibnamefont {Fiala}}, \bibinfo {author} {\bibfnamefont {J.}~\bibnamefont {Amato}}, \bibinfo {author} {\bibfnamefont {L.}~\bibnamefont {Tomáška}}, \bibinfo {author} {\bibfnamefont {J.}~\bibnamefont {Sponer}}, \bibinfo {author} {\bibfnamefont {J.}~\bibnamefont {Plavec}},\ and\ \bibinfo {author} {\bibfnamefont {L.}~\bibnamefont {Trantirek}},\ }\bibfield  {title} {\bibinfo {title} {Structure of a stable g-hairpin},\ }\href@noop {} {\bibfield  {journal} {\bibinfo  {journal} {Journal of the American Chemical Society}\ }\textbf {\bibinfo {volume} {139}},\ \bibinfo {pages} {3591} (\bibinfo {year} {2017})}\BibitemShut {NoStop}%
\bibitem [{\citenamefont {Bian}\ \emph {et~al.}(2018)\citenamefont {Bian}, \citenamefont {Ren}, \citenamefont {Song}, \citenamefont {Yu},\ and\ \citenamefont {Wang}}]{bian2018exploration}%
  \BibitemOpen
  \bibfield  {author} {\bibinfo {author} {\bibfnamefont {Y.}~\bibnamefont {Bian}}, \bibinfo {author} {\bibfnamefont {W.}~\bibnamefont {Ren}}, \bibinfo {author} {\bibfnamefont {F.}~\bibnamefont {Song}}, \bibinfo {author} {\bibfnamefont {J.}~\bibnamefont {Yu}},\ and\ \bibinfo {author} {\bibfnamefont {J.}~\bibnamefont {Wang}},\ }\bibfield  {title} {\bibinfo {title} {Exploration of the folding dynamics of human telomeric g-quadruplex with a hybrid atomistic structure-based model},\ }\href@noop {} {\bibfield  {journal} {\bibinfo  {journal} {The Journal of Chemical Physics}\ }\textbf {\bibinfo {volume} {148}} (\bibinfo {year} {2018})}\BibitemShut {NoStop}%
\bibitem [{\citenamefont {Bian}\ \emph {et~al.}(2020)\citenamefont {Bian}, \citenamefont {Song}, \citenamefont {Zhang}, \citenamefont {Yu}, \citenamefont {Wang},\ and\ \citenamefont {Wang}}]{bian2020insights}%
  \BibitemOpen
  \bibfield  {author} {\bibinfo {author} {\bibfnamefont {Y.}~\bibnamefont {Bian}}, \bibinfo {author} {\bibfnamefont {F.}~\bibnamefont {Song}}, \bibinfo {author} {\bibfnamefont {J.}~\bibnamefont {Zhang}}, \bibinfo {author} {\bibfnamefont {J.}~\bibnamefont {Yu}}, \bibinfo {author} {\bibfnamefont {J.}~\bibnamefont {Wang}},\ and\ \bibinfo {author} {\bibfnamefont {W.}~\bibnamefont {Wang}},\ }\bibfield  {title} {\bibinfo {title} {Insights into the kinetic partitioning folding dynamics of the human telomeric g-quadruplex from molecular simulations and machine learning},\ }\href@noop {} {\bibfield  {journal} {\bibinfo  {journal} {Journal of Chemical Theory and Computation}\ }\textbf {\bibinfo {volume} {16}},\ \bibinfo {pages} {5936} (\bibinfo {year} {2020})}\BibitemShut {NoStop}%
\bibitem [{\citenamefont {Stadlbauer}\ \emph {et~al.}(2021)\citenamefont {Stadlbauer}, \citenamefont {Islam}, \citenamefont {Otyepka}, \citenamefont {Chen}, \citenamefont {Monchaud}, \citenamefont {Zhou}, \citenamefont {Mergny},\ and\ \citenamefont {Sponer}}]{stadlbauer2021insights}%
  \BibitemOpen
  \bibfield  {author} {\bibinfo {author} {\bibfnamefont {P.}~\bibnamefont {Stadlbauer}}, \bibinfo {author} {\bibfnamefont {B.}~\bibnamefont {Islam}}, \bibinfo {author} {\bibfnamefont {M.}~\bibnamefont {Otyepka}}, \bibinfo {author} {\bibfnamefont {J.}~\bibnamefont {Chen}}, \bibinfo {author} {\bibfnamefont {D.}~\bibnamefont {Monchaud}}, \bibinfo {author} {\bibfnamefont {J.}~\bibnamefont {Zhou}}, \bibinfo {author} {\bibfnamefont {J.-L.}\ \bibnamefont {Mergny}},\ and\ \bibinfo {author} {\bibfnamefont {J.}~\bibnamefont {Sponer}},\ }\bibfield  {title} {\bibinfo {title} {Insights into g-quadruplex--hemin dynamics using atomistic simulations: implications for reactivity and folding},\ }\href@noop {} {\bibfield  {journal} {\bibinfo  {journal} {Journal of Chemical Theory and Computation}\ }\textbf {\bibinfo {volume} {17}},\ \bibinfo {pages} {1883} (\bibinfo {year} {2021})}\BibitemShut {NoStop}%
\bibitem [{\citenamefont {Maffeo}\ \emph {et~al.}(2012)\citenamefont {Maffeo}, \citenamefont {Luan},\ and\ \citenamefont {Aksimentiev}}]{Maffeo2012}%
  \BibitemOpen
  \bibfield  {author} {\bibinfo {author} {\bibfnamefont {C.}~\bibnamefont {Maffeo}}, \bibinfo {author} {\bibfnamefont {B.}~\bibnamefont {Luan}},\ and\ \bibinfo {author} {\bibfnamefont {A.}~\bibnamefont {Aksimentiev}},\ }\bibfield  {title} {\bibinfo {title} {End-to-end attraction of duplex dna},\ }\href@noop {} {\bibfield  {journal} {\bibinfo  {journal} {Nucleic Acids Res.}\ }\textbf {\bibinfo {volume} {40}},\ \bibinfo {pages} {3812} (\bibinfo {year} {2012})}\BibitemShut {NoStop}%
\bibitem [{\citenamefont {Maffeo}\ \emph {et~al.}(2014)\citenamefont {Maffeo}, \citenamefont {Yoo}, \citenamefont {Comer}, \citenamefont {Wells}, \citenamefont {Luan},\ and\ \citenamefont {Aksimentiev}}]{Maffeo2014}%
  \BibitemOpen
  \bibfield  {author} {\bibinfo {author} {\bibfnamefont {C.}~\bibnamefont {Maffeo}}, \bibinfo {author} {\bibfnamefont {J.}~\bibnamefont {Yoo}}, \bibinfo {author} {\bibfnamefont {J.}~\bibnamefont {Comer}}, \bibinfo {author} {\bibfnamefont {D.~B.}\ \bibnamefont {Wells}}, \bibinfo {author} {\bibfnamefont {B.}~\bibnamefont {Luan}},\ and\ \bibinfo {author} {\bibfnamefont {A.}~\bibnamefont {Aksimentiev}},\ }\bibfield  {title} {\bibinfo {title} {Close encounters with {DNA}},\ }\href {https://doi.org/10.1088/0953-8984/26/41/413101} {\bibfield  {journal} {\bibinfo  {journal} {J. Condens. Matter Phys.}\ }\textbf {\bibinfo {volume} {26}},\ \bibinfo {pages} {413101} (\bibinfo {year} {2014})}\BibitemShut {NoStop}%
\bibitem [{\citenamefont {Saurabh}\ \emph {et~al.}(2017)\citenamefont {Saurabh}, \citenamefont {Lansac}, \citenamefont {Jang}, \citenamefont {Glaser}, \citenamefont {Clark},\ and\ \citenamefont {Maiti}}]{GlaserPRE17}%
  \BibitemOpen
  \bibfield  {author} {\bibinfo {author} {\bibfnamefont {S.}~\bibnamefont {Saurabh}}, \bibinfo {author} {\bibfnamefont {Y.}~\bibnamefont {Lansac}}, \bibinfo {author} {\bibfnamefont {Y.~H.}\ \bibnamefont {Jang}}, \bibinfo {author} {\bibfnamefont {M.~A.}\ \bibnamefont {Glaser}}, \bibinfo {author} {\bibfnamefont {N.~A.}\ \bibnamefont {Clark}},\ and\ \bibinfo {author} {\bibfnamefont {P.~K.}\ \bibnamefont {Maiti}},\ }\bibfield  {title} {\bibinfo {title} {Understanding the origin of liquid crystal ordering of ultrashort double-stranded dna},\ }\href {https://doi.org/10.1103/PhysRevE.95.032702} {\bibfield  {journal} {\bibinfo  {journal} {Phys. Rev. E}\ }\textbf {\bibinfo {volume} {95}},\ \bibinfo {pages} {032702} (\bibinfo {year} {2017})}\BibitemShut {NoStop}%
\bibitem [{\citenamefont {Rosi}\ \emph {et~al.}(2023)\citenamefont {Rosi}, \citenamefont {Libera}, \citenamefont {Bertini}, \citenamefont {Orecchini}, \citenamefont {Corezzi}, \citenamefont {Schir{\`o}}, \citenamefont {Pernot}, \citenamefont {Biehl}, \citenamefont {Petrillo}, \citenamefont {Comez} \emph {et~al.}}]{rosi2023stacking}%
  \BibitemOpen
  \bibfield  {author} {\bibinfo {author} {\bibfnamefont {B.~P.}\ \bibnamefont {Rosi}}, \bibinfo {author} {\bibfnamefont {V.}~\bibnamefont {Libera}}, \bibinfo {author} {\bibfnamefont {L.}~\bibnamefont {Bertini}}, \bibinfo {author} {\bibfnamefont {A.}~\bibnamefont {Orecchini}}, \bibinfo {author} {\bibfnamefont {S.}~\bibnamefont {Corezzi}}, \bibinfo {author} {\bibfnamefont {G.}~\bibnamefont {Schir{\`o}}}, \bibinfo {author} {\bibfnamefont {P.}~\bibnamefont {Pernot}}, \bibinfo {author} {\bibfnamefont {R.}~\bibnamefont {Biehl}}, \bibinfo {author} {\bibfnamefont {C.}~\bibnamefont {Petrillo}}, \bibinfo {author} {\bibfnamefont {L.}~\bibnamefont {Comez}}, \emph {et~al.},\ }\bibfield  {title} {\bibinfo {title} {Stacking interactions and flexibility of human telomeric multimers},\ }\href@noop {} {\bibfield  {journal} {\bibinfo  {journal} {Journal of the American Chemical Society}\ }\textbf {\bibinfo {volume} {145}},\ \bibinfo {pages} {16166} (\bibinfo {year} {2023})}\BibitemShut {NoStop}%
\bibitem [{\citenamefont {Weik}\ \emph {et~al.}(2019)\citenamefont {Weik}, \citenamefont {Weeber}, \citenamefont {Szuttor}, \citenamefont {Breitsprecher}, \citenamefont {de~Graaf}, \citenamefont {Kuron}, \citenamefont {Landsgesell}, \citenamefont {Menke}, \citenamefont {Sean},\ and\ \citenamefont {Holm}}]{weik2019espresso}%
  \BibitemOpen
  \bibfield  {author} {\bibinfo {author} {\bibfnamefont {F.}~\bibnamefont {Weik}}, \bibinfo {author} {\bibfnamefont {R.}~\bibnamefont {Weeber}}, \bibinfo {author} {\bibfnamefont {K.}~\bibnamefont {Szuttor}}, \bibinfo {author} {\bibfnamefont {K.}~\bibnamefont {Breitsprecher}}, \bibinfo {author} {\bibfnamefont {J.}~\bibnamefont {de~Graaf}}, \bibinfo {author} {\bibfnamefont {M.}~\bibnamefont {Kuron}}, \bibinfo {author} {\bibfnamefont {J.}~\bibnamefont {Landsgesell}}, \bibinfo {author} {\bibfnamefont {H.}~\bibnamefont {Menke}}, \bibinfo {author} {\bibfnamefont {D.}~\bibnamefont {Sean}},\ and\ \bibinfo {author} {\bibfnamefont {C.}~\bibnamefont {Holm}},\ }\bibfield  {title} {\bibinfo {title} {Espresso 4.0--an extensible software package for simulating soft matter systems},\ }\href@noop {} {\bibfield  {journal} {\bibinfo  {journal} {The European Physical Journal Special Topics}\ }\textbf {\bibinfo {volume} {227}},\ \bibinfo {pages} {1789} (\bibinfo {year} {2019})}\BibitemShut {NoStop}%
\bibitem [{\citenamefont {Allen}\ and\ \citenamefont {Tildesley}(2017)}]{allen2017computer}%
  \BibitemOpen
  \bibfield  {author} {\bibinfo {author} {\bibfnamefont {M.~P.}\ \bibnamefont {Allen}}\ and\ \bibinfo {author} {\bibfnamefont {D.~J.}\ \bibnamefont {Tildesley}},\ }\href@noop {} {\emph {\bibinfo {title} {Computer simulation of liquids}}}\ (\bibinfo  {publisher} {Oxford university press},\ \bibinfo {year} {2017})\BibitemShut {NoStop}%
\bibitem [{\citenamefont {Weeks}\ \emph {et~al.}(1971)\citenamefont {Weeks}, \citenamefont {Chandler},\ and\ \citenamefont {Andersen}}]{weeks1971role}%
  \BibitemOpen
  \bibfield  {author} {\bibinfo {author} {\bibfnamefont {J.~D.}\ \bibnamefont {Weeks}}, \bibinfo {author} {\bibfnamefont {D.}~\bibnamefont {Chandler}},\ and\ \bibinfo {author} {\bibfnamefont {H.~C.}\ \bibnamefont {Andersen}},\ }\bibfield  {title} {\bibinfo {title} {Role of repulsive forces in determining the equilibrium structure of simple liquids},\ }\href@noop {} {\bibfield  {journal} {\bibinfo  {journal} {The Journal of chemical physics}\ }\textbf {\bibinfo {volume} {54}},\ \bibinfo {pages} {5237} (\bibinfo {year} {1971})}\BibitemShut {NoStop}%
\bibitem [{\citenamefont {Kremer}\ and\ \citenamefont {Grest}(1990)}]{kremer1990molecular}%
  \BibitemOpen
  \bibfield  {author} {\bibinfo {author} {\bibfnamefont {K.}~\bibnamefont {Kremer}}\ and\ \bibinfo {author} {\bibfnamefont {G.~S.}\ \bibnamefont {Grest}},\ }\bibfield  {title} {\bibinfo {title} {Molecular dynamics (md) simulations for polymers},\ }\href@noop {} {\bibfield  {journal} {\bibinfo  {journal} {Journal of Physics: Condensed Matter}\ }\textbf {\bibinfo {volume} {2}},\ \bibinfo {pages} {SA295} (\bibinfo {year} {1990})}\BibitemShut {NoStop}%
\bibitem [{\citenamefont {Libera}\ \emph {et~al.}(2021)\citenamefont {Libera}, \citenamefont {Andreeva}, \citenamefont {Martel}, \citenamefont {Thureau}, \citenamefont {Longo}, \citenamefont {Petrillo}, \citenamefont {Paciaroni}, \citenamefont {Schir{\`o}},\ and\ \citenamefont {Comez}}]{libera2021porphyrin}%
  \BibitemOpen
  \bibfield  {author} {\bibinfo {author} {\bibfnamefont {V.}~\bibnamefont {Libera}}, \bibinfo {author} {\bibfnamefont {E.~A.}\ \bibnamefont {Andreeva}}, \bibinfo {author} {\bibfnamefont {A.}~\bibnamefont {Martel}}, \bibinfo {author} {\bibfnamefont {A.}~\bibnamefont {Thureau}}, \bibinfo {author} {\bibfnamefont {M.}~\bibnamefont {Longo}}, \bibinfo {author} {\bibfnamefont {C.}~\bibnamefont {Petrillo}}, \bibinfo {author} {\bibfnamefont {A.}~\bibnamefont {Paciaroni}}, \bibinfo {author} {\bibfnamefont {G.}~\bibnamefont {Schir{\`o}}},\ and\ \bibinfo {author} {\bibfnamefont {L.}~\bibnamefont {Comez}},\ }\bibfield  {title} {\bibinfo {title} {Porphyrin binding and irradiation promote g-quadruplex dna dimeric structure},\ }\href@noop {} {\bibfield  {journal} {\bibinfo  {journal} {The Journal of Physical Chemistry Letters}\ }\textbf {\bibinfo {volume} {12}},\ \bibinfo {pages} {8096} (\bibinfo {year} {2021})}\BibitemShut {NoStop}%
\bibitem [{\citenamefont {Guinier}\ \emph {et~al.}(1955)\citenamefont {Guinier}, \citenamefont {Fournet}, \citenamefont {Walker},\ and\ \citenamefont {Yudowitch}}]{guinier1955small}%
  \BibitemOpen
  \bibfield  {author} {\bibinfo {author} {\bibfnamefont {A.}~\bibnamefont {Guinier}}, \bibinfo {author} {\bibfnamefont {G.}~\bibnamefont {Fournet}}, \bibinfo {author} {\bibfnamefont {C.~B.}\ \bibnamefont {Walker}},\ and\ \bibinfo {author} {\bibfnamefont {K.~L.}\ \bibnamefont {Yudowitch}},\ }\href@noop {} {\emph {\bibinfo {title} {Small-angle Scattering of X-rays}}}\ (\bibinfo  {publisher} {Wiley New York},\ \bibinfo {year} {1955})\BibitemShut {NoStop}%
\bibitem [{\citenamefont {Feigin}\ \emph {et~al.}(1987)\citenamefont {Feigin}, \citenamefont {Svergun} \emph {et~al.}}]{feigin1987structure}%
  \BibitemOpen
  \bibfield  {author} {\bibinfo {author} {\bibfnamefont {L.}~\bibnamefont {Feigin}}, \bibinfo {author} {\bibfnamefont {D.~I.}\ \bibnamefont {Svergun}}, \emph {et~al.},\ }\href@noop {} {\emph {\bibinfo {title} {Structure analysis by small-angle X-ray and neutron scattering}}},\ Vol.~\bibinfo {volume} {1}\ (\bibinfo  {publisher} {Springer},\ \bibinfo {year} {1987})\BibitemShut {NoStop}%
\bibitem [{\citenamefont {Bustamante}\ \emph {et~al.}(1994)\citenamefont {Bustamante}, \citenamefont {Marko}, \citenamefont {Siggia},\ and\ \citenamefont {Smith}}]{bustamante1994entropic}%
  \BibitemOpen
  \bibfield  {author} {\bibinfo {author} {\bibfnamefont {C.}~\bibnamefont {Bustamante}}, \bibinfo {author} {\bibfnamefont {J.~F.}\ \bibnamefont {Marko}}, \bibinfo {author} {\bibfnamefont {E.~D.}\ \bibnamefont {Siggia}},\ and\ \bibinfo {author} {\bibfnamefont {S.}~\bibnamefont {Smith}},\ }\bibfield  {title} {\bibinfo {title} {Entropic elasticity of $\lambda$-phage dna},\ }\href@noop {} {\bibfield  {journal} {\bibinfo  {journal} {Science}\ }\textbf {\bibinfo {volume} {265}},\ \bibinfo {pages} {1599} (\bibinfo {year} {1994})}\BibitemShut {NoStop}%
\bibitem [{\citenamefont {Beaucage}(1996)}]{beaucage1996small}%
  \BibitemOpen
  \bibfield  {author} {\bibinfo {author} {\bibfnamefont {G.}~\bibnamefont {Beaucage}},\ }\bibfield  {title} {\bibinfo {title} {Small-angle scattering from polymeric mass fractals of arbitrary mass-fractal dimension},\ }\href@noop {} {\bibfield  {journal} {\bibinfo  {journal} {Journal of applied crystallography}\ }\textbf {\bibinfo {volume} {29}},\ \bibinfo {pages} {134} (\bibinfo {year} {1996})}\BibitemShut {NoStop}%
\bibitem [{\citenamefont {Beaucage}(1995)}]{beaucage1995approximations}%
  \BibitemOpen
  \bibfield  {author} {\bibinfo {author} {\bibfnamefont {G.}~\bibnamefont {Beaucage}},\ }\bibfield  {title} {\bibinfo {title} {Approximations leading to a unified exponential/power-law approach to small-angle scattering},\ }\href@noop {} {\bibfield  {journal} {\bibinfo  {journal} {Journal of applied crystallography}\ }\textbf {\bibinfo {volume} {28}},\ \bibinfo {pages} {717} (\bibinfo {year} {1995})}\BibitemShut {NoStop}%
\bibitem [{\citenamefont {Sharp}(2016)}]{Sharp2016}%
  \BibitemOpen
  \bibfield  {author} {\bibinfo {author} {\bibfnamefont {K.~A.}\ \bibnamefont {Sharp}},\ }\bibfield  {title} {\bibinfo {title} {Unpacking the origins of in-cell crowding},\ }\href {https://doi.org/10.1073/pnas.1600098113} {\bibfield  {journal} {\bibinfo  {journal} {Proceedings of the National Academy of Sciences}\ }\textbf {\bibinfo {volume} {113}},\ \bibinfo {pages} {1684} (\bibinfo {year} {2016})},\ \Eprint {https://arxiv.org/abs/https://www.pnas.org/doi/pdf/10.1073/pnas.1600098113} {https://www.pnas.org/doi/pdf/10.1073/pnas.1600098113} \BibitemShut {NoStop}%
\bibitem [{\citenamefont {Rubinstein}\ and\ \citenamefont {Colby}(2003)}]{2003-rubinstein}%
  \BibitemOpen
  \bibfield  {author} {\bibinfo {author} {\bibfnamefont {M.}~\bibnamefont {Rubinstein}}\ and\ \bibinfo {author} {\bibfnamefont {R.~H.}\ \bibnamefont {Colby}},\ }\href@noop {} {\emph {\bibinfo {title} {Polymer Physics}}}\ (\bibinfo  {publisher} {Oxford University Press},\ \bibinfo {year} {2003})\BibitemShut {NoStop}%
\bibitem [{\citenamefont {Hsu}\ \emph {et~al.}(2010)\citenamefont {Hsu}, \citenamefont {Paul},\ and\ \citenamefont {Binder}}]{hsu2010standard}%
  \BibitemOpen
  \bibfield  {author} {\bibinfo {author} {\bibfnamefont {H.-P.}\ \bibnamefont {Hsu}}, \bibinfo {author} {\bibfnamefont {W.}~\bibnamefont {Paul}},\ and\ \bibinfo {author} {\bibfnamefont {K.}~\bibnamefont {Binder}},\ }\bibfield  {title} {\bibinfo {title} {Standard definitions of persistence length do not describe the local “intrinsic” stiffness of real polymer chains},\ }\href@noop {} {\bibfield  {journal} {\bibinfo  {journal} {Macromolecules}\ }\textbf {\bibinfo {volume} {43}},\ \bibinfo {pages} {3094} (\bibinfo {year} {2010})}\BibitemShut {NoStop}%
\bibitem [{\citenamefont {Sch{\"a}fer}\ and\ \citenamefont {Elsner}(2004)}]{schafer2004calculation}%
  \BibitemOpen
  \bibfield  {author} {\bibinfo {author} {\bibfnamefont {L.}~\bibnamefont {Sch{\"a}fer}}\ and\ \bibinfo {author} {\bibfnamefont {K.}~\bibnamefont {Elsner}},\ }\bibfield  {title} {\bibinfo {title} {Calculation of the persistence length of a flexible polymer chain with short-range self-repulsion},\ }\href@noop {} {\bibfield  {journal} {\bibinfo  {journal} {The European Physical Journal E}\ }\textbf {\bibinfo {volume} {13}},\ \bibinfo {pages} {225} (\bibinfo {year} {2004})}\BibitemShut {NoStop}%
\bibitem [{\citenamefont {Hsu}\ \emph {et~al.}(2013)\citenamefont {Hsu}, \citenamefont {Paul},\ and\ \citenamefont {Binder}}]{hsu2013estimation}%
  \BibitemOpen
  \bibfield  {author} {\bibinfo {author} {\bibfnamefont {H.-P.}\ \bibnamefont {Hsu}}, \bibinfo {author} {\bibfnamefont {W.}~\bibnamefont {Paul}},\ and\ \bibinfo {author} {\bibfnamefont {K.}~\bibnamefont {Binder}},\ }\bibfield  {title} {\bibinfo {title} {Estimation of persistence lengths of semiflexible polymers: Insight from simulations},\ }\href@noop {} {\bibfield  {journal} {\bibinfo  {journal} {Polymer Science Series C}\ }\textbf {\bibinfo {volume} {55}},\ \bibinfo {pages} {39} (\bibinfo {year} {2013})}\BibitemShut {NoStop}%
\bibitem [{\citenamefont {Forero-Martinez}\ \emph {et~al.}(2019)\citenamefont {Forero-Martinez}, \citenamefont {Baumeier},\ and\ \citenamefont {Kremer}}]{forero2019backbone}%
  \BibitemOpen
  \bibfield  {author} {\bibinfo {author} {\bibfnamefont {N.~C.}\ \bibnamefont {Forero-Martinez}}, \bibinfo {author} {\bibfnamefont {B.}~\bibnamefont {Baumeier}},\ and\ \bibinfo {author} {\bibfnamefont {K.}~\bibnamefont {Kremer}},\ }\bibfield  {title} {\bibinfo {title} {Backbone chemical composition and monomer sequence effects on phenylene polymer persistence lengths},\ }\href@noop {} {\bibfield  {journal} {\bibinfo  {journal} {Macromolecules}\ }\textbf {\bibinfo {volume} {52}},\ \bibinfo {pages} {5307} (\bibinfo {year} {2019})}\BibitemShut {NoStop}%
\bibitem [{\citenamefont {Micka}\ and\ \citenamefont {Kremer}(1996)}]{micka1996persistence}%
  \BibitemOpen
  \bibfield  {author} {\bibinfo {author} {\bibfnamefont {U.}~\bibnamefont {Micka}}\ and\ \bibinfo {author} {\bibfnamefont {K.}~\bibnamefont {Kremer}},\ }\bibfield  {title} {\bibinfo {title} {The persistence length of polyelectrolyte chains},\ }\href@noop {} {\bibfield  {journal} {\bibinfo  {journal} {Journal of Physics: Condensed Matter}\ }\textbf {\bibinfo {volume} {8}},\ \bibinfo {pages} {9463} (\bibinfo {year} {1996})}\BibitemShut {NoStop}%
\bibitem [{\citenamefont {Moyzis}\ \emph {et~al.}(1988)\citenamefont {Moyzis}, \citenamefont {Buckingham}, \citenamefont {Cram}, \citenamefont {Dani}, \citenamefont {Deaven}, \citenamefont {Jones}, \citenamefont {Meyne}, \citenamefont {Ratliff},\ and\ \citenamefont {Wu}}]{moyzis1988highly}%
  \BibitemOpen
  \bibfield  {author} {\bibinfo {author} {\bibfnamefont {R.~K.}\ \bibnamefont {Moyzis}}, \bibinfo {author} {\bibfnamefont {J.~M.}\ \bibnamefont {Buckingham}}, \bibinfo {author} {\bibfnamefont {L.~S.}\ \bibnamefont {Cram}}, \bibinfo {author} {\bibfnamefont {M.}~\bibnamefont {Dani}}, \bibinfo {author} {\bibfnamefont {L.~L.}\ \bibnamefont {Deaven}}, \bibinfo {author} {\bibfnamefont {M.~D.}\ \bibnamefont {Jones}}, \bibinfo {author} {\bibfnamefont {J.}~\bibnamefont {Meyne}}, \bibinfo {author} {\bibfnamefont {R.~L.}\ \bibnamefont {Ratliff}},\ and\ \bibinfo {author} {\bibfnamefont {J.-R.}\ \bibnamefont {Wu}},\ }\bibfield  {title} {\bibinfo {title} {A highly conserved repetitive dna sequence,(ttaggg) n, present at the telomeres of human chromosomes.},\ }\href@noop {} {\bibfield  {journal} {\bibinfo  {journal} {Proceedings of the National Academy of Sciences}\ }\textbf {\bibinfo {volume} {85}},\ \bibinfo {pages} {6622} (\bibinfo {year} {1988})}\BibitemShut {NoStop}%
\bibitem [{\citenamefont {Mostarac}\ \emph {et~al.}(2022)\citenamefont {Mostarac}, \citenamefont {Xiong}, \citenamefont {Gang},\ and\ \citenamefont {Kantorovich}}]{mostarac2022nanopolymers}%
  \BibitemOpen
  \bibfield  {author} {\bibinfo {author} {\bibfnamefont {D.}~\bibnamefont {Mostarac}}, \bibinfo {author} {\bibfnamefont {Y.}~\bibnamefont {Xiong}}, \bibinfo {author} {\bibfnamefont {O.}~\bibnamefont {Gang}},\ and\ \bibinfo {author} {\bibfnamefont {S.}~\bibnamefont {Kantorovich}},\ }\bibfield  {title} {\bibinfo {title} {Nanopolymers for magnetic applications: how to choose the architecture?},\ }\href@noop {} {\bibfield  {journal} {\bibinfo  {journal} {Nanoscale}\ }\textbf {\bibinfo {volume} {14}},\ \bibinfo {pages} {11139} (\bibinfo {year} {2022})}\BibitemShut {NoStop}%
\bibitem [{\citenamefont {Criscuolo}\ \emph {et~al.}(2022)\citenamefont {Criscuolo}, \citenamefont {Napolitano}, \citenamefont {Riccardi}, \citenamefont {Musumeci}, \citenamefont {Platella},\ and\ \citenamefont {Montesarchio}}]{criscuolo2022insights}%
  \BibitemOpen
  \bibfield  {author} {\bibinfo {author} {\bibfnamefont {A.}~\bibnamefont {Criscuolo}}, \bibinfo {author} {\bibfnamefont {E.}~\bibnamefont {Napolitano}}, \bibinfo {author} {\bibfnamefont {C.}~\bibnamefont {Riccardi}}, \bibinfo {author} {\bibfnamefont {D.}~\bibnamefont {Musumeci}}, \bibinfo {author} {\bibfnamefont {C.}~\bibnamefont {Platella}},\ and\ \bibinfo {author} {\bibfnamefont {D.}~\bibnamefont {Montesarchio}},\ }\bibfield  {title} {\bibinfo {title} {Insights into the small molecule targeting of biologically relevant g-quadruplexes: An overview of nmr and crystal structures},\ }\href@noop {} {\bibfield  {journal} {\bibinfo  {journal} {Pharmaceutics}\ }\textbf {\bibinfo {volume} {14}},\ \bibinfo {pages} {2361} (\bibinfo {year} {2022})}\BibitemShut {NoStop}%
\bibitem [{\citenamefont {Zegers}\ \emph {et~al.}(2023)\citenamefont {Zegers}, \citenamefont {Peters},\ and\ \citenamefont {Albada}}]{zegers2023dna}%
  \BibitemOpen
  \bibfield  {author} {\bibinfo {author} {\bibfnamefont {J.}~\bibnamefont {Zegers}}, \bibinfo {author} {\bibfnamefont {M.}~\bibnamefont {Peters}},\ and\ \bibinfo {author} {\bibfnamefont {B.}~\bibnamefont {Albada}},\ }\bibfield  {title} {\bibinfo {title} {Dna g-quadruplex-stabilizing metal complexes as anticancer drugs},\ }\href@noop {} {\bibfield  {journal} {\bibinfo  {journal} {JBIC Journal of Biological Inorganic Chemistry}\ }\textbf {\bibinfo {volume} {28}},\ \bibinfo {pages} {117} (\bibinfo {year} {2023})}\BibitemShut {NoStop}%
\end{thebibliography}%
\end{document}


\preprint{APS/123-QED}

\title{Supplementary Information \\
  \normalsize Polymeric Properties of Higher-Order G-Quadruplex Telomeric Structures: Effects of Chemically Inert Crowders}
\author{Deniz Mostarac}
\affiliation{%
Department of Physics, University of Rome La Sapienza, 00185 Rome, Italy}
\email{deniz.mostarac@uniroma1.it}

\author{Cristiano De Michele}
\affiliation{%
Department of Physics, University of Rome La Sapienza, 00185 Rome, Italy}%

\author{Mattia Trapella}
\affiliation{%
 Department of Physics and Geology, University of Perugia, 06123 Perugia, Italy}%

\author{Luca Bertini}
\affiliation{%
 Department of Physics and Geology, University of Perugia, 06123 Perugia, Italy}%

\author{Lucia Comez}
\affiliation{%
Department of Physics and Geology, University of Perugia, 06123 Perugia, Italy}%
 
\author{Alessandro Paciaroni}
\affiliation{%
 Department of Physics and Geology, University of Perugia, 06123 Perugia, Italy}%

\date{\today}
\maketitle

\section{Molecular dynamics}
Using the ESPResSo software package \cite{weik2019espresso}, we perform molecular dynamics simulations for G4 dimers, trimers and qudrimers at concentrations matching the experimental samples. The carrier fluid was represented implicitly, via the Langevin thermostat at fixed temperature $T$.\cite{allen2017computer} In practice it means that the Langevin equations of motion are integrated over time $t$ numerically:
\begin{equation}
    \label{eq:lang_trans}
    M_{i} \frac{d \vec{\mbox{$\nu$}}_{i}}{dt} = \vec{\mbox{F}}_{i} - \Gamma_{Tl} \vec{\mbox{$\nu$}}_{i} + 2 \vec{\mbox{$\xi$}}_{i}^{Tl},
\end{equation}
\begin{equation}
    \label{eq:lang_rot}
    I_{i} \frac{d \vec{\mbox{$\omega$}}_{i}}{dt} = \vec{\mbox{$\tau$}}_{i} - \Gamma_{R} \vec{\mbox{$\omega$}}_{i} + 2 \vec{\mbox{$\xi$}}_{i}^{R},
\end{equation}
\noindent where for the $i$-th particle in Eq. \eqref{eq:lang_trans},  $M_{i}$,is, in general, a rank two mass tensor, that in our case of isotropic monomers reduces to a scalar, $\vec{\mbox{F}}_{i}$ is the force acting on the particle,$\vec{\mbox{$\nu$}}_{i}$ denotes the translational velocity. $\Gamma_{Tl}$ denotes the translational friction tensor that once again in our particular case reduces to one scalar friction coefficient. Finally, $\vec{\mbox{$\xi$}}_{i}^{Tl}$ is a stochastic force, modelling the thermal fluctuations of the implicit solvent. Similarly, in Eq. \eqref{eq:lang_rot}, $I_{i}$ denotes $i$-th particle inertia tensor (scalar for a homogeneous sphere), $\vec{\mbox{$\tau$}}_{i}$ is torque acting on it, $\vec{\mbox{$\omega$}}_{i}$ is particle rotational velocity. As for the translation,  $\Gamma_{R}$ denotes the rotational friction tensor that reduces to a scalar for our monomers,  and the $\vec{\mbox{$\xi$}}_{i}^{R}$ is a stochastic torque serving for the same purpose as $\vec{\mbox{$\xi$}}_{i}^{Tl}$. Both stochastic terms satisfy the conditions on their time averages \cite{uhlenbeck1930theory}:
\begin{equation}
\langle \vec{\mbox{$\xi$}}^{Tl/R} \rangle_t = 0 \, ; 
\end{equation}
$$ \langle \vec{\mbox{$\xi$}}_{l}^{Tl/R}(t) \vec{\mbox{$\xi$}}_{k}^{Tl/R} (t^\prime) \rangle = 2\Gamma_{Tl/R} k_B T \delta_{l,k}\delta(t-t^\prime)  ;$$
where $k,l=x,y,z$.

Forces and torques in Eqs. \eqref{eq:lang_trans} and \eqref{eq:lang_rot} are calculated from inter-particle interaction potentials. Each simulation box contained 6000 G-tetrads at afixed concentration, that combine in to $6000/(3M)$ multimeres. We use periodic boundary conditions and a cubic simulation box to approximate infinite systems and be able to extract bulk properties at equilibrium. For the integration, the velocity Verlet algorithm was used,\cite{rapaport2004art} with a timestep of 0.01 (see \ref{ssec::units} for more detail on the simulation units). In all cases the initial configurations were generated so that both the positions and orientations of the largest predefined structures are appropriately randomised. We make sure sure that the system relaxes into an equilibrium configuration, by running an integration cycle for $2\cdot10^{6}$ integration steps. To obtain statistically significant results, we always present averages over eight independent simulation runs, 10 snapshots each, where the snapshots are sufficiently far apart from each other to minimise correlations (separated by $1\cdot10^{5}$ integrations). 
\section{Methodology}\label{sec:method}

The excluded volume of soft spheres with a characteristic diameter $\sigma$, is realised via the Weeks–Chandler–Andersen potential:
\cite{weeks1971role}:
\begin{equation}
    U_{WCA}(r)=
\begin{cases}
U_{LJ}(r)-U_{LJ}(r_{cut}),& \mathrm{if~} r<r_{cut}\\
 0 & \mathrm{otherwise}
\end{cases}
\label{eq:wca}  
\end{equation}
where and $U_{LJ}(r)$ is the conventional Lennard-Jones potential:
\begin{equation}
U_{LJ}(r)=4\epsilon \Biggl\{(\sigma/r)^{12}-(\sigma/r)^6\Biggr\}
\label{eq:LJ}
\end{equation}
where the cutoff value is $r_{cut}=2^{1/6}\sigma$. The parameter $\epsilon$ defines the interaction strength (relative to the energy scale). Only the center-of-mass (CoM) particle in each quartet is propagated using the equations of motion (Eqs.\ref{eq:lang_trans} and \ref{eq:lang_rot}, respectively). The rest of the soft sphere particles outlining the quartet are virtual, meaning that they have a fixed position w.r.t to the CoM particle, which incidentally is the only particle that carries mass. Note that the frictional coupling is set accordingly. The moment if inertia tensor of all CoM particles is modified to account for the halo of virtual sites outlining the quartet shape. Finitely extensible, non-linear elastic potential (FENE) used to model the linkers between the tetrads and G4 monomers has the following functional form:~\cite{kremer1990molecular}
\begin{equation}
 U_{FENE}(r) = -\frac{1}{2} K \Delta r_\mathrm{max}^2\ln \left[ 1 - \left(
         \frac{r-r_0}{\Delta r_\mathrm{max}} \right)^2 \right],
\label{eq_fene}
\end{equation}
where $K_f$ is the rigidity of the bond, $r_{f}$ is the maximal stretching length and $r_0$ is the equilibrium bond length. The virtual corner particles in adjacent sheets in a G4 are linked via the FENE bonds. Making multimeric structures (i.e. dimers, trimers and quadrimeres) out of G4 monomers is achieved by introducing FENE linkers between a randomly chosen pair of corner virtual particles on adjacent quartet sheets of neighbouring G4 monomers. The Lennard-Jones interaction (Eq.\ref{eq:LJ}), used to mimic the stacking interactions between monomers, is a good representation of the affinity monomers might have for the solvent and/or each other, and is often used in computational studies for this purpose \cite{coldstream2022gradient}. Tuning the stacking interaction is a simple but effective way to mimic the solvent in experiment, as long as one is exclusively interested in equilibrium properties. The CoM particles within the same G4 do not have a central attraction between them. 

\section{Reduced units and mapping to physical parameters}\label{ssec::units}

In this subsection we give a detailed overview of the units used in our simulations. We choose time scale and length scale in our MD simulations to be $[t]=1\times 10^{-9} \ s$ and $\sigma=0.4 nm$, respectively. Note that the length scale is equal to the diameter of a single particle in the $5\times 5$ grid of particles outlining the a G-tetrad. The value for $\sigma$ was chosen based on the values reported in \citet{rosi2023stacking} and \citet{libera2021porphyrin}. The energy scale in the simulations corresponds to the reduced temperature (in units of $k_B T$) of the Langevin thermostat,  which we set to $298.15 K$. This also corresponds to the steric repulsion strength $\epsilon_{WCA}$. The central attraction strength has been set to $\epsilon_{LJ}=5$. The above stated parameter choices uniquely define a mass scale. It is however completely arbitrary as far as the scope of this work is concerned. The factor $K$ of the potential given in Eq.~\eqref{eq_fene} is set to $K=10$. The equilibrium distance for the FENE bonds is set to $r_0=2\sigma$ while their maximum extension $r_{max}$, is set to $r_f=1.5r_0$. 
\section{Radius of Gyration}

\subsection{Guinier analysis}

\begin{table}[]
    \centering
    \begin{tabular}{|c|c|c|c|c|}
        \hline
        & $1\times G4$ & $2\times G4$ & $3\times G4$ & $4\times G4$ \\ 
        \hline
        $q_{min}$ $[\text{nm}^{-1}]$ & 0.1759 & 0.1331 & 0.1640 & 0.2016 \\ 
        $q_{max}$ $[\text{nm}^{-1}]$ & 1.2730 & 1.2103 & 1.2507 & 1.2769 \\ 
        $R_g$ $[\text{SI}]$ & 2.4449 & 4.2116 & 5.0006 & 6.1718 \\ 
        $R^2$ & 0.9640 & 0.9830 & 0.9933 & 0.9891 \\
        \hline
    \end{tabular}
    \caption{Guinier fit parameters on simulated scattering intensity profiles.}
    \label{tbl:guinier_sim}
\end{table}

\begin{table}[h!]
    \centering
    \begin{tabular}{|c|c|c|c|c|}
        \hline
        & $2JSL$ & $TEL48$ & $TEL72$ & $TEL96$ \\ 
        \hline
        $q_{min}$ $[\text{nm}^{-1}]$ & 0.1738 & 0.1260 & 0.1734 & 0.2143 \\ 
        $q_{max}$ $[\text{nm}^{-1}]$ & 1.2879 & 1.1837 & 1.2791 & 1.2948 \\ 
        $R_g$ $[\text{SI}]$ & 2.5805 & 3.9858 & 5.2875 & 6.5606 \\ 
        $R^2$ & 0.9734 & 0.9943 & 0.9962 & 0.9946 \\
        \hline
    \end{tabular}
    \caption{Guinier fit parameters on the experimental scattering intensity data from \citet{monsen2021solution}}
    \label{tbl:guinier_exp}
\end{table}

We wrote a small function that iteratively solves for $R_g$ adjusting the $q_{min}$ and $q_{max}$ so that coefficient of determination $R^2$ is maximised, with the added constraint that $q_{max}<1.3$. $R_g$ is extracted via the standard procedure where scattering intensity is considered as a function of $(qR_g)^2$, where in the Guinier region, it should appear as a linear function on a log-linear plot. Therefore ,we can extract both the $R_g$ and $R^2$ parameters form linear regression in the appropriate range. Final fit parameters are provided in Tables~\ref{tbl:guinier_sim} and \ref{tbl:guinier_exp}, for simulated and experimental data, respectively. 

\subsection{Direct calculation from simulated data}
\begin{table}[t!]
    \centering
    
    \begin{tabular}{|c|c|c|c|c|}
        \hline
        & $1\times G4$ & $2\times G4$ & $3\times G4$ & $4\times G4$ \\ 
        \hline
        $R_g$ $[\text{SI}]$ & $2.519 \pm 0.0029$ & $4.2 \pm 0.022$ & $5.47\pm 0.018$ & $6.57 \pm 0.0914$ \\ 
        \hline
    \end{tabular}
    \caption{$R_g$ calculated directly from simulated data using Eq.\ref{eq:rg_direct}}
    \label{tbl:rg_direct}
\end{table}
In Table~\ref{tbl:rg_direct}, we provide $R_g$ values and standard deviation, as calculated directly using:
\begin{equation}\label{eq:rg_direct}
    R_g = \sqrt{\lambda_1^2+\lambda_2^2+\lambda_3^2}
\end{equation}
, where $\lambda_1>\lambda_2>\lambda_3$ are the eigenvalues of the gyration tensor:
\begin{align}
    G_{\mu \nu}=\dfrac{1}{N}\sum^N_{i=1}(r_{i,\mu}-r_{cm,\mu})(r_{i,\nu}-r_{cm,\nu})
\end{align}
where $r_{i,\mu}$ and $r_{cm,\mu}$ are the $\mu$-th Cartesian components of the position of the $i$-th particle and the center of mass, respectively. The summation is carried over all $N$ particles.




\bibliography{bibliography_quadriplex,magfils} 